\documentclass[a4paper]{article}
\usepackage{graphicx} % Required for inserting images
\usepackage{amsmath}
\usepackage{amssymb}
\usepackage[round]{natbib}
\usepackage{xcolor}
\usepackage{authblk}
\usepackage{url}

\title{Dealing with partial missing correlations in multivariate and surrogate meta-analyses}
\author[1]{Riccardo De Santis\thanks{Corresponding address: riccardo.desantis@unipd.it}}
\author[1]{Annamaria Guolo}
\affil[1]{University of Padova, Department of Statistical Sciences, Italy}
\date{\today}

\begin{document}

\maketitle

\begin{abstract}
    This work addresses the issue of partially missing correlations within the framework of bivariate and surrogate meta-analyses. While restricting the analysis to complete-case studies may appear to constitute the most straightforward analytical strategy, such an approach has been demonstrated to yield substantial inefficiencies and potential bias in the resulting estimates. Current methodological contributions in the literature circumvent this limitation either through aggregate estimation procedures grounded in likelihood-based frameworks under simplifying assumptions, or by resorting to deterministic imputation strategies, such as the empirical mean derived from observed units. In the present paper, we propose a multiple imputation framework in which imputation is performed via stochastic procedures based on a Beta regression model, thereby explicitly accounting for the missing at random (MAR) assumption underlying the observed missingness mechanism. We demonstrate the effectiveness of our method through an extensive simulation study across various scenarios, comparing our proposal with simple mean imputation and complete-case analysis under the MAR assumption.
\end{abstract}

\section{Introduction}
In recent decades, the development of multivariate meta-analysis techniques has been suggested in the literature to jointly synthesize multiple outcomes obtained by different studies, while properly accounting for their correlation; see, e.g., \cite{jackson2011multivariate} and \cite[Chapter~9]{schmid2020handbook}. Common examples include the meta-analysis for the evaluation of the accuracy of diagnostic tests using study-specific measures of specificity and sensitivity \citep{ma2016statistical} and the meta-analysis of the evaluation of clinical endpoints and surrogate markers \citep{collier2023handling}. 

As noted by \cite[pp.164]{schmid2020handbook}, conducting a series of univariate meta-analyses on each outcome entails substantial methodological limitations, particularly in the context of joint inference, whose statistical efficiency is inherently compromised by the failure to account for the correlation structure among outcomes. Contributions to multivariate modeling were first introduced by \cite{raudenbush1988educational, van2002advanced, gleser2009stochastically}. The formal extension of the two-stage linear random-effects meta-analysis model was subsequently established by \cite{jackson2010extending, jackson2011multivariate}, under the assumption of joint normality of the outcomes. The incorporation of study-specific covariates was further addressed in \cite{gasparrini2012multivariate} and in \cite{sera2019extended}.

The random-effects formulation of the multivariate meta-analytic model accounts for two distinct sources of correlation among the outcomes, or main effects: the within-study correlation at the study level and the between-study correlation, the latter reflecting the relationships among the true underlying study-specific effects. An alternative parameterization of the bivariate model can be used to describe the evaluation of surrogate endpoints, used as an alternative measure of clinical outcomes, usually obtained earlier, more often, or with less invasive methods. The definition of surrogate was first formally introduced in \cite{prentice1989surrogate}. According to this formulation, the between-study model has the main objective of characterizing the (linear) relationship between the clinical outcome and the surrogate endpoint. This asymmetric formulation was first formalized in a frequentist statistical model by \cite{daniels1997meta}, and closely related proposals are the model specifications of \cite{buyse2000validation, korn2005assessing}. More recent Bayesian approaches can be found in \cite{bujkiewicz2016bayesian, collier2023handling}, while the works of \cite{bujkiewicz2017uncertainty, vickers2025comparison} focus on comparisons of the performances of the existing model proposed in the literature.
%% missing correlations
When working with multivariate meta-analysis, one practical issue is the possible lack of information about the within-study correlations, a topic that has received considerable attention in the literature; see, e.g., \cite{ishak2008impact, riley2007bivariate, riley2009multivariate, jackson2020multivariate}. Partial or complete missingness of within-study correlations is often addressed with simple solutions. In some approaches, missing correlations are simply ignored, as discussed, for example, in \cite{korn2005assessing, thompson2005can, ishak2008impact, collier2023handling}. However, such a solution can impact inferential conclusions \citep{riley2009multivariate}. When individual patient data are not available to estimate within-study correlations, they can be replaced by some fixed values. For example, \cite{daniels1997meta} propose the use of a common value based on prior knowledge or on the sample average of known correlations when the missingness is partial. In special situations, missing correlations can be obtained by explicit formulae. See the results in \cite{trikalinos2008method} for mutually exclusive binary outcomes and \cite{wei2013estimating} for related binary outcomes. 
%for approximate ad hoc formulae to estimate the correlation for logarithmic odds ratios or standardized mean differences, when individual data are not reported. 
 Eventually, these approaches rely on the missing completely at random (MCAR) assumption. \cite{collier2023handling} suggest a Bayesian methodology to deal with partial missingness, where a transformation of the correlations, e.g., the Fisher transformation, is assumed to have a normal prior distribution. It allows for the more challenging missing at random (MAR) mechanism, but it is not the main focus of their simulations, and further some relevant limits in terms of unbiasedness of the posteriors are present. On the other hand, \cite{riley2008alternative} propose a model in the case of complete missingness and bivariate outcome, through a meta-analysis model in which the within-correlations and the between-study correlation are collapsed in a unique parameter capturing all the correlation effects, which allows model estimation and inference on the main effects with no computational problems. However, this approach is not suitable when the correlation structure of the underlying model is of primary interest.

This paper suggests a new methodology to handle partial missingness of within-study correlations, allowing for the MAR mechanism. The proposal is a multiple imputation procedure based on beta regression \citep{kieschnick2003regression, ferrari2004beta}, which preserves the natural properties of correlations. In this sense, we adopt the generalization of the Beta distribution over the interval $[-1,1]$. Compared to the normal linear regression approach with the Fisher transform of observed observations, our approach remains valid even when observed values are close to the boundaries; see, e.g., \cite{bishara2012testing}. 
Furthermore, well-established results in the literature \citep{hotelling1953new} show that a generalization of the beta distribution is the natural distribution of the sample correlation coefficients of normally distributed outcomes, with strong connections to the distribution of sample variance matrices, the Wishart distribution \citep{olkin1964multivariate, khatri1965some}.
The choice of a multiple imputation procedure, over a simpler deterministic univariate imputation, is known to reflect the additional randomness more accurately in the final model estimation \citep{vanbuuren2018flexible}. Our imputation method can be interpreted as partially Bayesian in the spirit of \cite{rubinmultiple1987}, since we first draw the regression coefficients from their normal distribution and then use them to sample the imputed values from the beta distribution. Although the framework allows for general multivariate outcomes, in this paper we will focus on two-dimensional outcomes. Two main reasons are well highlighted in \cite{jackson2011multivariate}. First, multivariate modeling is not without challenges, the most notable being that multivariate meta-analyses can lead to estimation difficulties, due to the more complex relationships involved and the larger number of unknown parameters. Second, and most relevantly for the purposes of the present manuscript, within-study correlations are often not reported in individual studies. Both issues are further exacerbated in higher dimensions. A simulation study is conducted to evaluate the proposed multiple imputation procedure under different scenarios, assuming a missing at random mechanism. The scenarios are divided into two main blocks: the bivariate model and the surrogate model. The proposed method is compared with complete-case analysis and sample mean imputation. The results of the full-data model are also reported as a benchmark for completeness.

The remainder of the paper is organized as follows. Section \ref{sec:metamodel} introduces the meta-analytic model of interest, Section \ref{sec:imput} details the imputation strategy, Section \ref{sec:sims} presents the simulation study under different scenarios, Section \ref{sec::realdata} contains a real data application, and Section \ref{sec:discuss} provides some concluding remarks.

\section{Bivariate and surrogate model specification}\label{sec:metamodel}
Following \cite{jackson2010extending, jackson2011multivariate}, suppose that the multivariate meta-analysis is based on $K$ studies, each of them providing a bivariate outcome
\begin{equation*}
    y_i=\begin{pmatrix}
        y_{i1} \\ y_{i2}
    \end{pmatrix}, \ \ i=1,\dots,K.
\end{equation*}
As in the univariate meta-analysis (e.g., \cite{stijnen2020analysis}), the random-effects formulation of the model considers the within-study model relating the outcome to the random effects $\theta_i=(\theta_{i1},\theta_{i2})^T$,
\begin{equation}\label{eqn:model_within}
    y_i|\theta_i \sim \mathcal{N} 
    \begin{pmatrix}
    \theta_i,
    S_i
    \end{pmatrix},
\end{equation}
where
%\begin{equation*}
   % \theta_i=
 %  \begin{pmatrix}
  %      \theta_{i1} \\
  %      \theta_{i2}
  %  \end{pmatrix},
%\end{equation*}
\begin{equation*}
S_i =
\begin{pmatrix}
\sigma_{i1}^2 & \rho_{w,i} \sigma_{i1} \sigma_{i2} \\
\rho_{w,i} \sigma_{i1} \sigma_{i2} & \sigma_{i2}^2
\end{pmatrix}
\end{equation*}
and $\rho_{w,i}$ represents the within-study correlation between $y_{i1}$ and $y_{i2}$. Between-study variances $\sigma_{i1}$ and $\sigma_{i2}$ of $y_{i1}$ and $y_{i2}$, respectively, are usually considered as known and equal to the sample variance estimated from the studies included in the meta-analysis. This is common practice in the univariate case, provided that each study has a sufficiently large sample size.
The between-study model specifies a normal distribution for $\theta_i$, namely,
\begin{equation*}
\theta_i  \sim \mathcal{N}
\begin{pmatrix}
    \mu,T
\end{pmatrix},
\end{equation*}
where
\begin{equation*}
    \mu=
    \begin{pmatrix}
        \mu_1 \\ \mu_2
    \end{pmatrix}\ \ {\rm and} \ \ 
    T=
    \begin{pmatrix}
    \tau_1^2 & \rho_b \tau_1 \tau_2 \\
    \rho_b \tau_1 \tau_2 & \tau_2^2
    \end{pmatrix},
\end{equation*}
where $\rho_{b}$ represents the between-study correlation and $\tau^2_1$ and $\tau^2$ are the between-study variances of the random effects $\theta_{i1}$ and $\theta_{i2}$, respectively.
According to the specifications above, it turns out that, marginally,
\begin{equation}\label{eqn:model_marginal}
y_i \sim \mathcal{N}(\mu, S_i + T).
\end{equation}
Alternative within-study models that avoid normal approximations and are based on the original data from each study included in the meta-analysis can be adopted, although at the price of an increased computational burden. See, for example, the multivariate models used in the meta-analysis for the accuracy of diagnostic tests that involve binomial distributions for the within-study models \citep{chen2017simple, liu2020meta}.  

As stated in the Introduction, within single studies the estimates $r_{w,i}$ of the within-study correlations $\rho_{w,i}$ are often unavailable or not reported; consequently, the estimation of $\rho_b$ is commonly bypassed in practice. A notable example is provided in \cite{riley2008alternative}, who propose to overcome the problem by creating a bivariate model with a single correlation parameter that jointly captures the within-study and between-study correlations. Despite its computational feasibility, this approach presents non-trivial interpretation issues and does not constitute an adequate solution in settings where the correlation structure of the model is of primary interest. Nevertheless, there are instances where $\rho_b$ is of direct inferential interest, or where multivariate two-stage modeling is essential to prevent systematic bias in the estimation of target parameters. An example is given by \textit{surrogate endpoint models}, which are constructed to examine the association between clinical endpoints and surrogate endpoints in the context of randomized controlled trials \citep{daniels1997meta, collier2023handling}. Here, the bivariate outcome comprises a target quantity, e.g., a treatment effect on the clinical endpoint, and a surrogate quantity, the latter being of interest insofar as it approximates the target outcome while offering the practical advantage of being more readily obtainable in clinical practice. \cite{collier2023handling} describes the bivariate random-effects model for a surrogate related to the clinical outcome as in (\ref{eqn:model_marginal}), or, alternatively, with a parameterization that makes a linear relationship between the outcome and the surrogate explicit. Let $\theta_{1i}$ be the clinical outcome and let $\theta_{2i}$ be the surrogate endpoint of study $i$, which can be obtained earlier, more often or with less invasive procedures than $\theta_{1i}$. A useful reparametrization of the between-study model leads to
\begin{equation*}
    \theta_{1i}|\theta_{2i} \sim (\delta_0+\delta_1 \theta_{2i}, \sigma^2_e), \ \theta_{2i} \sim N(\mu_s, \sigma^2_s),
\end{equation*}
where the primary inferential focus is on the target parameter $\delta_1$, which represents the degree of adherence of the surrogate outcome to the clinical endpoint, thus providing a quantitative measure of its validity as a surrogate marker. The parameter $\sigma^2_e$ represents the between-study heterogeneity, which should ideally be small in order to have a good prediction of the outcome. The within-study model resembles (\ref{eqn:model_within}). 
As \cite{collier2023handling} point out, the within-study correlations of a surrogate endpoint model are rarely reported, and their evaluation would require access to individual patient data. The literature typically ignores the presence of within-study correlations, i.e. fixes the within-study correlations to zero \citep{korn2005assessing}, or substitutes them with a fixed value. In the seminal paper by \cite{daniels1997meta}, the missing within-study correlations are suggested to be set at 0.2 or at a smaller value, using a sensitivity analysis to choose among alternatives, if necessary. Such solutions have been shown to lead to biased inference that can lead to an unreliable assessment of surrogate quality, as \cite{collier2023handling} illustrate. Simulation results supporting these conclusions are also reported in the Simulation Study Section. An alternative solution to deal with missing within-study correlations based on multiple imputation is described in Section~\ref{sec:imput}.

\section{Imputation strategy using Beta regression}\label{sec:imput}
This paper investigates the application of multiple imputation strategies \citep{rubinmultiple1987} to address the problem of missing within-study correlations in multivariate meta-analysis. The imputation procedure is performed using Beta regression \citep{ferrari2004beta}, which represents a natural and powerful modeling framework for the regression analysis of bounded outcomes. Let $r_i, i \in {1,\dots,K}$ denote the observed correlation taking values in the interval $[-1, 1]$. We assume that each $r_i$ is a random variable that arises from the linear transformation $r_i = -1 + 2 r_{i}^*$, where $r_{i}^*$ follows a Beta distribution with parameters $(\mu_i, \phi)$ under the mean-precision parametrization.
Following standard regression modeling conventions, we assume that the conditional expected value is expressed as a function of observed covariates
$$
g(r_i)= X_i^T \gamma,
$$
where $g(\cdot)$ denotes the link function to be specified, $X_{i}, \; i \in \{1,\dots, K \}$ is a $p$-dimensional vector of observed study-level covariates with the parameter vector $\gamma$. Usually, we expect $p$ to be low in a meta-analysis framework, but we do not make any peculiar restriction. Since variance is a function of both $\mu_i$ and $\phi$, the model naturally accommodates heteroscedasticity. Regarding the mean submodel, although several choices for the link function are admissible, the most widely adopted specification \citep{ferrari2004beta} is the logit transformation $g(r_i^*)=\log \{r_i^*/(1-r_i^*)\}$. Model estimation is performed using standard maximum likelihood, closely related to the framework of generalized linear models \citep{agresti2015foundations}.
To develop the multiple imputation strategy, let $\mathcal{S}$ and $\mathcal{N}$ denote the index sets of units with, respectively, observed and missing correlation coefficients. Furthermore, let $X_j, j \in {1,\dots,n}$ denote the covariates observed in all studies. Following \cite{rubinmultiple1987} and \cite{vanbuuren2018flexible}, the proposed algorithm proceeds according to the following steps.
\begin{enumerate}
\item Transform the observed correlation coefficients $r_i, i \in \mathcal{S}$ into the interval $[0,1]$;
\item Fit a Beta regression model using all observations in $\mathcal{S}$;
\item Repeat $M$ times:
\begin{enumerate}
\item Draw a sample for the regression coefficients $(\gamma, \phi)$ from their multivariate normal distribution;
\item For each $j \in \mathcal{N}$, compute the shape parameters $(\alpha_j, \beta_j)$ of the Beta distribution as a function of $X_j$;
\item For each $j \in \mathcal{N}$, draw a sample from a Beta distribution with the corresponding estimated parameters;
\item Reverse the imputed values to the interval $[-1,1]$;
\item Fit the meta-analytic model of interest.
\end{enumerate}
\item Aggregate the $M$ results according to Rubin's rules.
\end{enumerate}
The aggregation of multiple imputation estimates is performed using Rubin's rules \citep{rubinmultiple1987}. For any target parameter $\psi$, the final estimate is obtained as the arithmetic mean
\begin{equation*}
\overline{\psi}=\frac{\sum_{m=1}^M\hat{\psi}_{m}}{M},
\end{equation*}
whose associated variance estimate
\begin{equation*}
\hat{U}=\overline{V}_M+(1+M^{-1})B_M.
\end{equation*}
decomposes into two components;
\begin{equation*}
\overline{V}_M=\sum_{m=1}^M \hat{V}_m/M ; \quad B_M=\frac{\sum_{m=1}^M(V_m-\overline{V}_M)^2}{M-1}.
\end{equation*}
The reference distribution for inference is a Student's $t$ with $\nu=\left\{ \left(M-1\right)\left(1+r^{-1}_M\right) \right\}$  degrees of freedom, where
$$
r_M=(1+M^{-1})B/\overline{V}_M.
$$
As discussed in \cite{vanbuuren2018flexible}, for practical purposes setting the total number of imputations to $M=5$ is generally sufficient to ensure reliability while limiting the influence of Monte Carlo variability inherent to the multiple imputation procedure.

\section{Simulation study}\label{sec:sims}
A simulation study was conducted to evaluate the performance of the multiple imputation method introduced in Section \ref{sec:imput} against competing approaches. The competing methods considered are complete case analysis and sample mean imputation; results obtained under full data availability are also reported as a benchmark. Section \ref{subsec:sims1} examines the impact of missing within-study correlations in the bivariate meta-analytic model, while Section \ref{subsec:sims2} investigates the surrogate endpoint model. The \texttt{R} code to perform the multiple imputation method proposed in Section \ref{sec:imput} can be found on GitHub \url{https://github.com/Rickdesa/Imputing-correlation-coefficients}.

\subsection{Bivariate model}\label{subsec:sims1}
Data are simulated for the meta-analytic model described in Section \ref{sec:metamodel} under different scenarios, for a total of $5,000$ replicates per case. The number $K$ of studies is set to values in $\{20, 30\}$. The vector of the true mean effects $(\mu_1, \mu_2)^T$ is set equal to $(1,0)^T$; the between-study variances $\tau^2_{1}$ and $\tau^2_{2}$ are equal to $1$ and $0.5$, respectively, while the between-study correlation $\rho_b$ takes values in $\{-0.3,0.0,0.3,0.5,0.8\}$. The within-study variances are drawn from a Beta distribution with parameters $(1.5,4.0)$, so as to yield a median close to $0.25$, as in \cite{riley2007bivariate} and \cite{riley2009multivariate}. The generation of within-study correlations is more involved. Starting from a standard normal covariate $X$, the within-study correlations are simulated from a Beta distribution with mean equal to $\mbox{expit}\{0.5X + \eta\}$ and dispersion parameter $\phi$; the resulting values are then rescaled to the interval $[-1, 1]$. The parameters $(\eta, \phi)^T$ are set equal to $(2.20, 3.2)^T$, yielding an unconditional mean of $0.78$ and variance of $0.1$; to $(1.16, 9.6)^T$, yielding a mean of $0.5$ and variance of $0.1$; to $(0.54, 16.5)^T$, yielding a mean of $0.25$ and variance of $0.1$; and to $(0, 20)^T$, yielding a mean of $0$ and variance of $0.1$. Missing data are generated through an MAR mechanism depending on the covariate $X$ defined above, with the average proportion of missing values taking values in $\{0.3, 0.5\}$. The missingness model is specified as a logistic regression model with parameter $\gamma=0.6$, with the intercept adapted to achieve the desired average proportion of missing values. Finally, the bivariate meta-analytic model is fitted to the reconstructed data using the \texttt{R} package \textit{mvmeta} (\url{https://cran.r-project.org/web/packages/mvmeta/index.html}, \cite{gasparrini2012multivariate}). For the multiple imputation procedure, we use $M=5$ replicates, following \cite{vanbuuren2018flexible}.

\begin{figure}
    \centering
 \includegraphics[width=4.8in]{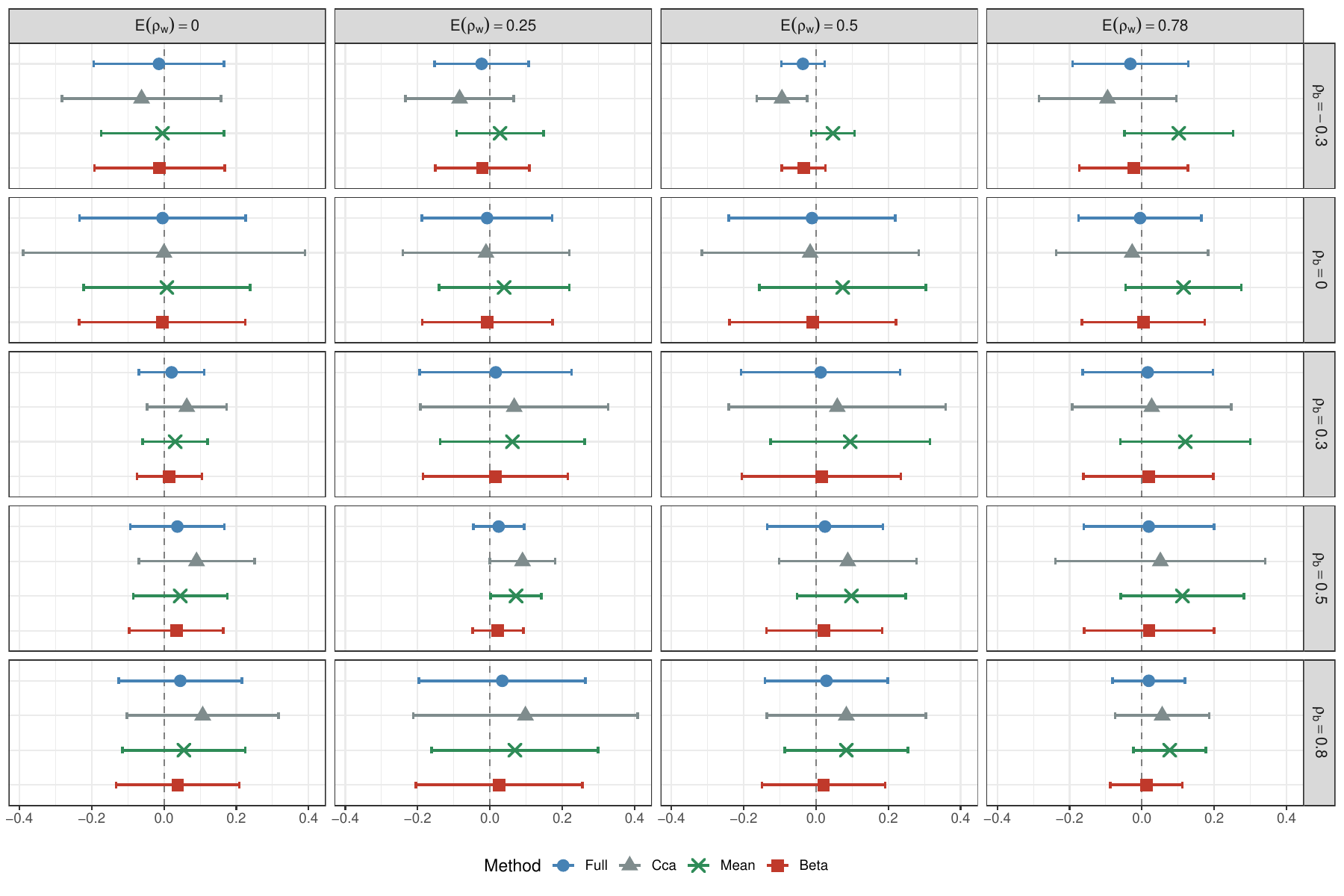}
    \caption{Median bias and median absolute deviation (MAD) of the estimator of the correlation coefficient $\rho_b$ in the bivariate model, under different mean of the within-study correlation (columns) and between-study correlation (rows); sample size $K=20$, average proportion of rejection 50\%. Methods: bivariate model with fully available data (Full), complete case analysis (Cca), sample average imputation of missing $\rho_w$ (Mean), stochastic multiple imputation procedure for missing $\rho_w$ (Beta).}
    \label{fig:bivar1}
\end{figure}

The simulation results are reported in terms of median bias and median absolute deviation of the parameter of interest given by the between-study correlation $\rho_b$. Figure \ref{fig:bivar1} reports the median bias and median absolute deviation for the parameter $\rho_b$, for $K=20$. The use of a complete-case analysis gives results with median bias that increases as the between-study correlation and the within-study correlation differ substantially. Imputing the within-study correlation with the sample mean does not improve upon the results, as the median bias increases with the true within-study correlations, regardless of the set-up value of $\rho_b$. In addition, the median bias of the sample mean imputation method tends to be positive.
A slight improvement over the complete-case analysis is obtained in the variability given by the median absolute deviation, which is consistently lower. The suggested multiple imputation approach, conversely, consistently yields estimates of $\rho_b$ close to those from the target full data model, in terms of both bias and variability. The performance is maintained regardless of the simulation value of $\rho_b$, $\rho_W$ and sample size $K$. See the results for $K=30$ in the Supplementary Material, Table 3. For sake of completeness, we report also the numeric data of Figure \ref{fig:bivar1} in Table 2 of the Supplementary Material.

A similar analysis for an average missing data proportion equal to $30\%$ (Tables 4--5) is reported in the Supplementary Material. The results show that the median bias of both the sample mean imputation and the complete-case analysis is slightly attenuated compared to the case with $50\%$ of missing data, yet remains non-negligible.
Finally, the Supplementary Material also includes the simulation results for the estimators of the main effects $(\theta_1, \theta_2)$ (Tables 6--13) and the true sampling correlation coefficient between the estimators (Tables 14--17), showing that none of the analyzed procedures show relevant bias across all scenarios.

\subsection{Surrogate model}\label{subsec:sims2}
Again, $5,000$ replicates are run for each parameter configuration. The number of studies is set to $K \in \{20, 30\}$. The vector of the true mean effects $(\mu_1, \mu_2)^T$ is set equal to $(1, 0)^T$; the between-study variances $\tau^2_1$ and $\tau^2_2$ are equal to $1$ and $0.5$, respectively, while the between-study correlation $\rho_b$ takes values in $\{-0.4, 0, 0.4, 0.7, 0.9\}$. The within-study variances are drawn from a Beta distribution with shape parameters $(2, 7)$. The generation of within-study correlations is more complex. Starting from a standard normal covariate $X$, the within-study correlations are simulated from a Beta distribution with mean equal to $\mbox{expit}{(0.5X + \eta)}$ and dispersion parameter $\phi$; the resulting values are then rescaled to the interval $[-1, 1]$. The parameters $(\eta, \phi)^T$ are set equal to $(2.20, 3.2)^T$, yielding an unconditional mean of $0.78$ and variance of $0.1$; to $(1.16, 9.6)^T$, yielding a mean of $0.5$ and variance of $0.1$; to $(0.54, 16.5)^T$, yielding a mean of $0.25$ and variance of $0.1$; and to $(0, 20)^T$, yielding a mean of $0$ and variance of $0.1$. Missing data are generated through an MAR mechanism depending on the covariate $X$ defined above, with an average missing data proportion of $0.5$. The results for an average missing data proportion of $0.3$ are provided in the Supplementary Material. The missingness model is specified as a logistic regression model with parameter $\gamma=0.6$, with the intercept adapted to achieve the desired average proportion of missing values. Finally, the surrogate meta-analytic model is fitted to the reconstructed data using a standard maximum likelihood approach. The results are reported in terms of median bias and median absolute deviation of the parameters of interest. For the multiple imputation procedure, $M=5$ replicates are used, following \cite{vanbuuren2018flexible}.

\begin{figure}
    \centering
 \includegraphics[width=4.8in]{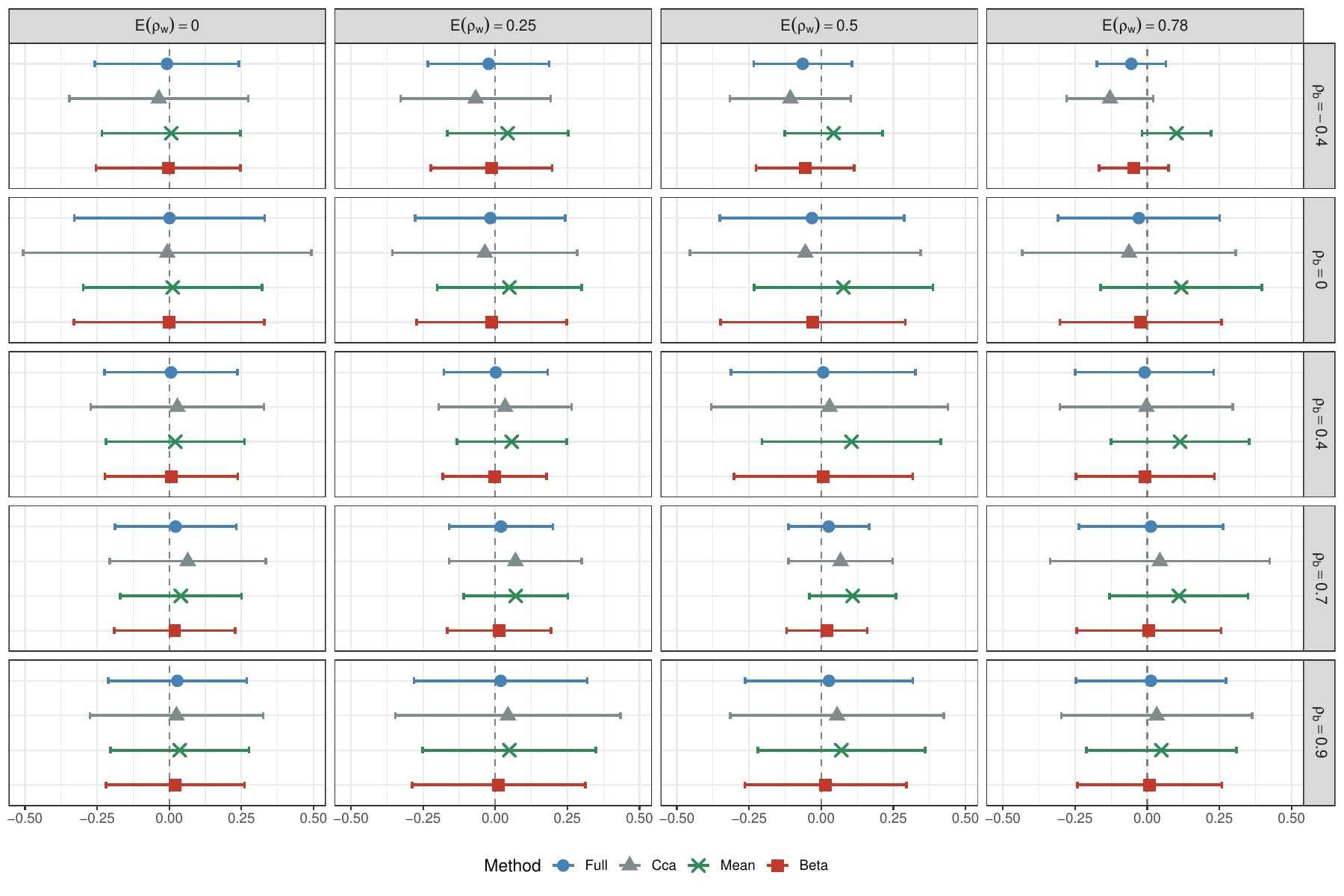}
    \caption{Median bias and median absolute deviation (MAD) of the estimator of the regression parameter $\delta_1$ in the surrogate model, under different mean of the within-study correlation (columns) and between-study correlation (rows); sample size $K=20$, average proportion of rejection 50\%. Methods: bivariate model with fully available data (Full), complete case analysis (Cca), sample average imputation of missing $\rho_w$ (Mean), stochastic multiple imputation procedure for missing $\rho_w$ (Beta).}
    \label{fig:surrog1}
\end{figure}

Figure \ref{fig:surrog1} reports the simulation results for inference on the regression coefficient $\delta_1$, for $K=20$ and $50\%$ average missing data. Similarly to the previous case, using the sample average imputation leads to positively biased estimates of the regression coefficient $\delta_1$, especially for large values of the within-study correlation. The complete-case analysis tends to suffer from bias in the estimators when the between-study and within-study correlation coefficients differ substantially. In addition, this approach is characterized by larger estimated standard errors than the other alternatives.
The proposed Beta imputation solution has a satisfactory performance, with very small bias for the estimator of $\delta_1$, consistently across all parameter configurations. The results are very close to the target given by the full data analysis. Table 19 of the Supplementary Material reports the same analysis for $K=30$, where it can be seen that the full data analysis and the proposed imputation method perform well in the limiting cases mentioned above. For sake of completeness, we report also the numeric data of Figure \ref{fig:surrog1} in Table 18 of the Supplementary Material.

The results for an average missing data proportion of $30\%$ are reported in the Supplementary Material (Tables 20--21), where it can be noted that the median bias remains substantial even at this lower proportion of missing data. Furthermore, the Supplementary Material (Tables 22--25) also reports results for the residual variance $\sigma^2_e$. A notable negative bias is observed, particularly in the smaller sample size scenario; this is somewhat expected, and a restricted maximum likelihood approach may be more appropriate when this parameter is of primary interest. However, the purpose of the present manuscript is to highlight the close agreement between the proposed method and the complete-data model.

\section{Real data application}\label{sec::realdata}
As a numerical illustration, we use the data from a meta-analysis for the evaluation of potential surrogate markers in \cite{daniels1997meta}. The original dataset contains 24 studies measuring outcomes related to a test treatment and a standard treatment. In particular, each study included in the meta-analysis contains information about the clinical outcome for patients, represented by the development of AIDS or death over 2 years, and the surrogate marker, given by the difference in mean change in CD4 cell count between baseline and 6 months, for studies of the AIDS Clinical Trial Group. As the CD4 cell count decreases, the risk of various opportunistic infections and malignancies increases, which is indicative of AIDS diagnosis and death. Data in \cite{daniels1997meta} include the estimated standard errors and sample correlation coefficients associated with the point estimates of the clinical outcome and the potential surrogate marker.
The sample correlation coefficients are highly concentrated, ranging from -0.22 to 0.17, with mean -0.0996 and standard error 0.0839.

Starting from the full data information, we create 50\% of missing correlation coefficients through a logit model. In particular, we create a dichotomous variable based on whether the test treatment is a dose treatment of a particular class
of anti-HIV, represented by 600 mg of zidovudine (ZDV[600]), with regression coefficients equal to 0.6. For the new dataset with partially missing correlations all the methods described in  Section \ref{sec:sims} are used to estimate the parameter $\delta_1$ in Section \ref{sec:metamodel} useful to understand the association between the clinical outcome and the surrogate. Table (\ref{tab:example}) reports the results in terms of estimate of $\delta_1$ and associated standard error. While the differences among the methods are quite subtle, the proposed Beta imputation approach provides the results closest to the full data analysis.
\begin{table}[htbp]
\centering
\label{tab:example}
\begin{tabular}{c |c c c c}
\hline
Method & Full & Cca & Mean & Beta \\ \hline
Estimates & -0.01152 & -0.00900 & -0.01135 & -0.01142  \\ 
Standard errors & 0.00421 & 0.00558 & 0.00422 & 0.00421 \\
\hline
\end{tabular}
\caption{Real data application based on the data in \cite{daniels1997meta}. Estimates of the parameter $\delta_1$ are reported. Methods: bivariate model with fully available data (Full), complete case analysis (Cca), sample average imputation of missing $\rho_w$ (Mean), stochastic multiple imputation procedure for missing $\rho_w$ (Beta).}
\end{table}

\section{Discussion}\label{sec:discuss}
In this manuscript, we propose a way to handle partially missing correlations in a bivariate meta-analysis model with correlated outcomes, which includes the bivariate model for potential surrogates as a special case. The proposal is a multiple imputation approach, where imputation is performed using a Beta regression model that can account for the missing at random assumption underlying the missingness mechanism. Simulation results under different scenarios, used to compare the proposed method to classical solutions in the literature, have shown satisfactory performance when, on average, half of the within-study correlations are missing. The proposed method is easily implementable and requires minimal computational effort, providing a valid alternative to existing solutions for applied researchers.

Starting from the encouraging results, an interesting extension of the work could go in the direction of three or higher dimensional outcomes, or multiple surrogates. In such cases, the main limitation lies in the full model, which generally struggles when the number of studies available is small, as is often the case.

\section{Acknowledgements}
The authors Riccardo De Santis and Annamaria Guolo acknowledge the following funding. The publication was produced with funding from the Italian Ministry of University and Research as part of the Call for Proposals for the scrolling of the final rankings of the PRIN 2022 - Project title ”MEMIMR: Measurement Errors and Missing Information in Meta-Regression” - Project No.2022FZY9PM - CUP C53C24000740006.

\bibliographystyle{plainnat} 
\bibliography{bibtex.bib}

\newpage

%\documentclass[a4paper]{article}
%\usepackage{graphicx} % Required for inserting images
%\usepackage{amsmath}
%\usepackage{amssymb}
%\usepackage[round]{natbib}
%\usepackage{xcolor}
%\usepackage{url}
%\title{Supplementary material: Dealing with partial missing correlations in multivariate meta-analysis}
%\author{Riccardo De Santis, Annamaria Guolo}
%\date{}

%\begin{document}

%\maketitle

\section{Supplementary material: Dealing with partial missing correlations in multivariate meta-analysis}

\subsection{Bivariate model analysis}
\begin{table}[!htbp] \centering 
	\caption{Median bias (median absolute deviation) of the correlation coefficient $\rho_b$, $K=20$ in the bivariate model, average proportion of rejection 50\%. Full stands for the model with fully available data, Cca for complete case analysis, Mean for the sample average imputation of missing $\rho_w$, and Beta for the stochastic multiple imputation procedure for missing $\rho_w$.} 
 \label{tab:bivar1} 
\begin{tabular}{@{\extracolsep{5pt}} cccccc} 
\\[-1.8ex]\hline 
\hline \\[-1.8ex] 
$\rho_b$ & $\mathbb{E}[\rho_w]$ & Full & Cca & Mean & Beta \\ 
\hline \\[-1.8ex] 
-0.3 & 0 & -0.014 (0.18) & -0.062 (0.22) & -0.004 (0.17) &  -0.012 (0.18) \\ 
-0.3 & 0.25 & -0.023 (0.13) & -0.084 (0.15) & 0.028 (0.12) &  -0.021 (0.13) \\ 
-0.3 & 0.50 & -0.036 (0.06) & -0.094 (0.07) & 0.047 (0.06) &  -0.034 (0.06) \\ 
-0.3 & 0.78 & -0.032 (0.16) & -0.095 (0.19) & 0.102 (0.15) &  -0.023 (0.15) \\ 
0 & 0 & -0.004 (0.23) & -0.000 (0.39) & 0.008 (0.23) & -0.005 (0.23) \\ 
0 & 0.25 & -0.008 (0.18) & -0.011 (0.23) & 0.039 (0.18) &  -0.007 (0.18) \\ 
0 & 0.50 & -0.011 (0.23) & -0.016 (0.30) & 0.074 (0.23) &  -0.009 (0.23) \\ 
0 & 0.78 & -0.005 (0.17) & -0.027 (0.21) & 0.115 (0.16) &  0.004 (0.17) \\ 
0.3 & 0 & 0.021 (0.09) & 0.063 (0.11) & 0.031 (0.09) &  0.015 (0.09) \\ 
0.3 & 0.25 & 0.016 (0.21) & 0.067 (0.26) & 0.062 (0.20) &  0.015 (0.20) \\ 
0.3 & 0.50 & 0.013 (0.22) & 0.059 (0.30) & 0.095 (0.22) &  0.015 (0.22) \\ 
0.3 & 0.78 & 0.016 (0.18) & 0.027 (0.22) & 0.120 (0.18) &  0.018 (0.18) \\ 
0.5 & 0 & 0.037 (0.13) & 0.090 (0.16) & 0.045 (0.13) &  0.034 (0.13) \\ 
0.5 & 0.25 & 0.024 (0.07) & 0.090 (0.09) & 0.072 (0.07) &  0.022 (0.07) \\ 
0.5 & 0.50 & 0.025 (0.16) & 0.088 (0.19) & 0.098 (0.15) &  0.023 (0.16) \\ 
0.5 & 0.78 & 0.019 (0.18) & 0.051 (0.29) & 0.112 (0.17) &  0.020 (0.18) \\ 
0.8 & 0 & 0.045 (0.17) & 0.107 (0.21) & 0.055 (0.17)  & 0.038 (0.17) \\ 
0.8 & 0.25 & 0.034 (0.23) & 0.098 (0.31) & 0.069 (0.23) & 0.025 (0.23) \\ 
0.8 & 0.50 & 0.029 (0.17) & 0.084 (0.22) & 0.084 (0.17) &  0.021 (0.17) \\ 
0.8 & 0.78 & 0.019 (0.10) & 0.056 (0.13) & 0.077 (0.10)  & 0.012 (0.10) \\ 
\hline \\[-1.8ex] 
\end{tabular} 
\end{table} 

\begin{table}[!htbp] \centering 
	\caption{Median bias (median absolute deviation) of the correlation coefficient $\rho_b$ in the bivariate model, $K=30$, average proportion of rejection 50\%.  Full stands for the model with fully available data, Cca for complete case analysis, Mean for the sample average imputation of missing $\rho_w$, and Beta for the stochastic multiple imputation procedure for missing $\rho_w$.} 
 \label{tab:bivar2} 
\begin{tabular}{@{\extracolsep{5pt}} cccccc} 
\\[-1.8ex]\hline 
\hline \\[-1.8ex] 
$\rho_b$ & $\mathbb{E}[\rho_w]$ & Full & Cca & Mean  & Beta \\ 
\hline \\[-1.8ex] 
-0.3 & 0 & -0.021 (0.20) & -0.042 (0.26) & -0.009 (0.20) & -0.020 (0.20) \\ 
-0.3 & 0.25 & -0.023 (0.23) & -0.046 (0.29) & 0.022 (0.22) &  -0.020 (0.22) \\ 
-0.3 & 0.50 & -0.025 (0.18) & -0.051 (0.23) & 0.060 (0.18) & -0.021 (0.18) \\ 
-0.3 & 0.78 & -0.024 (0.14) & -0.051 (0.17) & 0.099 (0.13) & -0.016 (0.14) \\ 
0 & 0 & -0.008 (0.08) & -0.001 (0.10) & 0.004 (0.08)  & -0.008 (0.08) \\ 
0 & 0.25 & -0.003 (0.17) & 0.003 (0.20) & 0.043 (0.16)  & -0.002 (0.16) \\ 
0 & 0.50 & -0.012 (0.13) & -0.015 (0.19) & 0.070 (0.12) & -0.011 (0.13) \\ 
0 & 0.78 & -0.013 (0.17) & -0.031 (0.21) & 0.106 (0.17) &  -0.008 (0.17) \\ 
0.3 & 0 & 0.008 (0.23) & 0.038 (0.29) & 0.022 (0.23) & 0.007 (0.23) \\ 
0.3 & 0.25 & 0.015 (0.18) & 0.037 (0.23) & 0.058 (0.18) & 0.012 (0.18) \\ 
0.3 & 0.50 & 0.003 (0.11) & 0.014 (0.13) & 0.078 (0.12)  & -0.001 (0.11) \\ 
0.3 & 0.78 & 0.001 (0.21) & -0.002 (0.26) & 0.104 (0.21)  & 0.004 (0.21) \\ 
0.5 & 0 & 0.021 (0.21) & 0.060 (0.27) & 0.035 (0.21) & 0.018 (0.21) \\ 
0.5 & 0.25 & 0.022 (0.18) & 0.056 (0.23) & 0.067 (0.18) & 0.019 (0.18) \\ 
0.5 & 0.50 & 0.010 (0.15) & 0.033 (0.18) & 0.080 (0.15) & 0.005 (0.15) \\ 
0.5 & 0.78 & 0.009 (0.09) & 0.026 (0.12) & 0.100 (0.09)  & 0.008 (0.09) \\ 
0.8 & 0 & 0.028 (0.17) & 0.070 (0.21) & 0.039 (0.17) & 0.022 (0.17) \\ 
0.8 & 0.25 & 0.021 (0.06) & 0.059 (0.09) & 0.057 (0.06) & 0.017 (0.07) \\ 
0.8 & 0.50 & 0.014 (0.17) & 0.043 (0.21) & 0.067 (0.17)  & 0.005 (0.17) \\ 
0.8 & 0.78 & 0.010 (0.16) & 0.027 (0.25) & 0.063 (0.15)  & 0.003 (0.16) \\ 
\hline \\[-1.8ex] 
\end{tabular} 
\end{table}

 \begin{table}[!htbp] \centering 
 	\caption{Median bias (median absolute deviation) of the correlation coefficient $\rho_b$ in the bivariate model, $K=20$, average proportion of rejection 30\%.  Full stands for the model with fully available data, Cca for complete case analysis, Mean for the sample average imputation of missing $\rho_w$, and Beta for the stochastic multiple imputation procedure for missing $\rho_w$.} 
  \label{tab:bivar_app_1} 
\begin{tabular}{@{\extracolsep{5pt}} cccccc} 
\\[-1.8ex]\hline 
\hline \\[-1.8ex] 
$\rho_b$ & $\mathbb{E}[\rho_w]$ & Full & Cca & Mean  & Beta \\ 
\hline \\[-1.8ex] 
-0.3 & 0 & -0.015 (0.23) & -0.039 (0.39) & -0.008 (0.22) &  -0.016 (0.23) \\ 
-0.3 & 0.25 & -0.021 (0.18) & -0.041 (0.29) & 0.011 (0.18) & -0.019 (0.18) \\ 
-0.3 & 0.50 & -0.036 (0.23) & -0.053 (0.39) & 0.021 (0.23) &  -0.033 (0.23) \\ 
-0.3 & 0.78 & -0.033 (0.17) & -0.051 (0.26) & 0.053 (0.16) &  -0.027 (0.17) \\ 
0 & 0 & -0.003 (0.09) & -0.006 (0.12) & 0.007 (0.08) &  -0.004 (0.09) \\ 
0 & 0.25 & -0.006 (0.21) & -0.008 (0.33) & 0.024 (0.20) &  -0.007 (0.21) \\ 
0 & 0.50 & -0.010 (0.22) & -0.011 (0.38) & 0.046 (0.22) &  -0.008 (0.22) \\ 
0 & 0.78 & -0.006 (0.18) & -0.011 (0.28) & 0.068 (0.17) &  -0.003 (0.18) \\ 
0.3 & 0 & 0.021 (0.13) & 0.035 (0.21) & 0.029 (0.13) & 0.018 (0.13) \\ 
0.3 & 0.25 & 0.017 (0.07) & 0.034 (0.11) & 0.047 (0.07) &  0.016 (0.07) \\ 
0.3 & 0.50 & 0.013 (0.16) & 0.023 (0.25) & 0.066 (0.15) &  0.014 (0.16) \\ 
0.3 & 0.78 & 0.015 (0.16) & 0.021 (0.25) & 0.081 (0.14) & 0.014 (0.16) \\ 
0.5 & 0 & 0.037 (0.17) & 0.049 (0.27) & 0.044 (0.17) &  0.034 (0.17) \\ 
0.5 & 0.25 & 0.025 (0.23) & 0.053 (0.40) & 0.055 (0.23) &  0.025 (0.23) \\ 
0.5 & 0.50 & 0.025 (0.17) & 0.045 (0.29) & 0.072 (0.17) &  0.025 (0.17) \\ 
0.5 & 0.78 & 0.018 (0.10) & 0.032 (0.10) & 0.073 (0.10) &  0.017 (0.10) \\ 
0.8 & 0 & 0.045 (0.20) & 0.065 (0.35) & 0.052 (0.20) &  0.040 (0.20) \\ 
0.8 & 0.25 & 0.035 (0.22) & 0.058 (0.37) & 0.057 (0.23) &  0.028 (0.22) \\ 
0.8 & 0.50 & 0.029 (0.18) & 0.050 (0.28) & 0.061 (0.18) & 0.026 (0.18) \\ 
0.8 & 0.78 & 0.018 (0.14) & 0.035 (0.21) & 0.051 (0.13) & 0.015 (0.14) \\ 
\hline \\[-1.8ex] 
\end{tabular} 
\end{table}

\begin{table}[!htbp] \centering 
  \caption{Median bias (median absolute deviation) of the correlation coefficient $\rho_b$ in the bivariate model, $K=30$, average proportion of rejection 30\%. Full stands for the model with fully available data, Cca for complete case analysis, Mean for the sample average imputation of missing $\rho_w$, and Beta for the stochastic multiple imputation procedure for missing $\rho_w$.} 
  \label{tab:bivar_app_2} 
\begin{tabular}{@{\extracolsep{5pt}} cccccc} 
\\[-1.8ex]\hline 
\hline \\[-1.8ex] 
$\rho_b$ & $\mathbb{E}[\rho_w]$ & Full & Cca & Mean & Beta \\ 
\hline \\[-1.8ex] 
-0.3 & 0 & -0.021 (0.08) & -0.028 (0.12) & -0.012 (0.08) &  -0.020 (0.08) \\ 
-0.3 & 0.25 & -0.023 (0.17) & -0.039 (0.25) & 0.010 (0.16) &  -0.021 (0.16) \\ 
-0.3 & 0.50 & -0.025 (0.07) & -0.037 (0.11) & 0.032 (0.07) &  -0.022 (0.08) \\ 
-0.3 & 0.78 & -0.024 (0.17) & -0.036 (0.26) & 0.055 (0.18) &  -0.021 (0.17) \\ 
0 & 0 & -0.008 (0.23) & -0.002 (0.39) & -0.002 (0.23) &  -0.007 (0.23) \\ 
0 & 0.25 & -0.003 (0.18) & -0.012 (0.31) & 0.026 (0.18) &  -0.005 (0.18) \\ 
0 & 0.50 & -0.012 (0.12) & -0.016 (0.09) & 0.040 (0.12) &  -0.013 (0.11) \\ 
0 & 0.78 & -0.013 (0.21) & -0.014 (0.36) & 0.060 (0.21) &  -0.011 (0.21) \\ 
0.3 & 0 & 0.008 (0.21) & 0.017 (0.35) & 0.017 (0.21) &  0.008 (0.21) \\ 
0.3 & 0.25 & 0.015 (0.18) & 0.022 (0.29) & 0.043 (0.19)  & 0.013 (0.18) \\ 
0.3 & 0.50 & 0.003 (0.15) & 0.007 (0.23) & 0.052 (0.15)  & 0.002 (0.15) \\ 
0.3 & 0.78 & 0.001 (0.09) & 0.010 (0.13) & 0.066 (0.09)  & 0.001 (0.09) \\ 
0.5 & 0 & 0.021 (0.17) & 0.038 (0.28) & 0.031 (0.17) & 0.020 (0.17) \\ 
0.5 & 0.25 & 0.022 (0.20) & 0.034 (0.32) & 0.050 (0.18) &  0.021 (0.20) \\ 
0.5 & 0.50 & 0.010 (0.17) & 0.020 (0.26) & 0.052 (0.17) &  0.008 (0.17) \\ 
0.5 & 0.78 & 0.009 (0.23) & 0.021 (0.30) & 0.063 (0.23) &  0.007 (0.23) \\ 
0.8 & 0 & 0.028 (0.16) & 0.043 (0.20) & 0.036 (0.15) &  0.024 (0.16) \\ 
0.8 & 0.25 & 0.021 (0.07) & 0.036 (0.09) & 0.044 (0.07) &  0.017 (0.08) \\ 
0.8 & 0.50 & 0.014 (0.20) & 0.025 (0.25) & 0.044 (0.19) &  0.007 (0.20) \\ 
0.8 & 0.78 & 0.010 (0.23) & 0.018 (0.29) & 0.042 (0.23) &  0.008 (0.23) \\ 
\hline \\[-1.8ex] 
\end{tabular} 
\end{table}

\begin{table}[!htbp] \centering 
  \caption{Median bias (median absolute deviation) of the fixed effect $\theta_1$ in the bivariate model, $K=20$, average proportion of rejection 50\%. Full stands for the model with fully available data, Cca for complete case analysis, Mean for the sample average imputation of missing $\rho_w$, and Beta for the stochastic multiple imputation procedure for missing $\rho_w$.} 
  %\label{} 
\begin{tabular}{@{\extracolsep{5pt}} cccccc} 
\\[-1.8ex]\hline 
\hline \\[-1.8ex] 
$\rho_b$ & $\mathbb{E}[\rho_w]$ & Full & Cca & Mean & Beta \\ 
\hline \\[-1.8ex] 
-0.3 & 0 & 0.004 (0.14) & -0.005 (0.16) & 0.004 (0.14) &  0.005 (0.14) \\ 
-0.3 & 0.25 & -0.001 (0.14) & -0.006 (0.16) & -0.003 (0.14) &  -0.001 (0.14) \\ 
-0.3 & 0.50 & 0.001 (0.13) & 0.003 (0.16) & -0.001 (0.13)  & 0.001 (0.13) \\ 
-0.3 & 0.78 & 0.001 (0.14) & -0.006 (0.16) & -0.002 (0.14)  & 0.000 (0.14) \\ 
0 & 0 & 0.004 (0.17) & -0.003 (0.24) & 0.003 (0.17)  & 0.004 (0.17) \\ 
0 & 0.25 & -0.000 (0.14) & -0.006 (0.16) & -0.002 (0.14) &  -0.002 (0.14) \\ 
0 & 0.50 & -0.000 (0.17) & 0.003 (0.20) & -0.001 (0.17) &  -0.000 (0.17) \\ 
0 & 0.78 & -0.001 (0.17) & -0.004 (0.20) & 0.002 (0.17) &  -0.002 (0.17) \\ 
0.3 & 0 & 0.004 (0.17) & -0.006 (0.20) & 0.003 (0.17)  & 0.004 (0.17) \\ 
0.3 & 0.25 & -0.002 (0.17) & -0.007 (0.20) & -0.001 (0.17) &  -0.001 (0.17) \\ 
0.3 & 0.50 & -0.003 (0.16) & 0.005 (0.20) & -0.000 (0.17) &  -0.001 (0.17) \\ 
0.3 & 0.78 & -0.003 (0.14) & 0.002 (0.17) & -0.003 (0.14) & -0.003 (0.14) \\ 
0.5 & 0 & 0.005 (0.14) & -0.005 (0.16) & 0.004 (0.14) &  0.006 (0.14) \\ 
0.5 & 0.25 & -0.001 (0.14) & -0.006 (0.16) & -0.002 (0.14) &  -0.000 (0.14) \\ 
0.5 & 0.50 & -0.002 (0.14) & 0.005 (0.17) & -0.000 (0.14) &  -0.002 (0.14) \\ 
0.5 & 0.78 & -0.003 (0.14) & 0.001 (0.19) & -0.003 (0.14) &  -0.003 (0.14) \\ 
0.8 & 0 & 0.004 (0.14) & -0.004 (0.17) & 0.005 (0.14) &  0.006 (0.14) \\ 
0.8 & 0.25 & 0.001 (0.17) & -0.003 (0.20) & 0.001 (0.16)  & 0.001 (0.17) \\ 
0.8 & 0.50 & -0.001 (0.17) & 0.002 (0.20) & -0.001 (0.17)  & 0.000 (0.17) \\ 
0.8 & 0.78 & -0.002 (0.17) & -0.004 (0.20) & 0.000 (0.17)  & -0.002 (0.17) \\ 
\hline \\[-1.8ex] 
\end{tabular} 
\end{table} 

\begin{table}[!htbp] \centering 
  \caption{Median bias (median absolute deviation) of the fixed effect $\theta_1$ in the bivariate model, $K=30$, average proportion of rejection 50\%. Full stands for the model with fully available data, Cca for complete case analysis, Mean for the sample average imputation of missing $\rho_w$, and Beta for the stochastic multiple imputation procedure for missing $\rho_w$.} 
  %\label{} 
\begin{tabular}{@{\extracolsep{5pt}} cccccc} 
\\[-1.8ex]\hline 
\hline \\[-1.8ex] 
$\rho_b$ & $\mathbb{E}[\rho_w]$ & Full & Cca & Mean & Beta \\  
\hline \\[-1.8ex] 
-0.3 & 0 & 0.005 (0.17) & -0.001 (0.20) & 0.002 (0.17) & 0.003 (0.17) \\ 
-0.3 & 0.25 & 0.007 (0.17) & 0.007 (0.20) & 0.007 (0.17) & 0.007 (0.17) \\ 
-0.3 & 0.50 & 0.002 (0.14) & 0.003 (0.16) & 0.001 (0.14) & 0.001 (0.14) \\ 
-0.3 & 0.78 & -0.000 (0.14) & 0.007 (0.16) & 0.001 (0.14) & -0.000 (0.14) \\ 
0 & 0 & 0.000 (0.14) & 0.004 (0.16) & 0.001 (0.14) &  0.002 (0.14) \\ 
0 & 0.25 & 0.006 (0.14) & 0.008 (0.16) & 0.005 (0.14) &  0.005 (0.14) \\ 
0 & 0.50 & -0.001 (0.14) & 0.005 (0.19) & -0.000 (0.14) & 0.001 (0.14) \\ 
0 & 0.78 & -0.002 (0.14) & 0.006 (0.17) & -0.000 (0.14) & -0.003 (0.14) \\ 
0.3 & 0 & 0.003 (0.17) & 0.005 (0.20) & 0.004 (0.17) &  0.003 (0.17) \\ 
0.3 & 0.25 & 0.002 (0.17) & 0.005 (0.20) & 0.005 (0.17) &  0.003 (0.17) \\ 
0.3 & 0.50 & -0.002 (0.17) & 0.003 (0.20) & -0.002 (0.17)  & -0.002 (0.17) \\ 
0.3 & 0.78 & -0.005 (0.17) & 0.001 (0.21) & -0.002 (0.17) & -0.003 (0.17) \\ 
0.5 & 0 & 0.004 (0.17) & 0.004 (0.20) & 0.005 (0.17) &  0.004 (0.16) \\ 
0.5 & 0.25 & 0.003 (0.14) & 0.006 (0.17) & 0.006 (0.14) &  0.004 (0.14) \\ 
0.5 & 0.50 & -0.001 (0.13) & 0.003 (0.16) & -0.001 (0.13)  & -0.001 (0.13) \\ 
0.5 & 0.78 & -0.004 (0.13) & 0.001 (0.16) & -0.004 (0.14) &  -0.004 (0.14) \\ 
0.8 & 0 & 0.002 (0.14) & 0.006 (0.16) & 0.005 (0.14) &  0.003 (0.14) \\ 
0.8 & 0.25 & 0.003 (0.13) & 0.007 (0.19) & 0.005 (0.14) &  0.003 (0.13) \\ 
0.8 & 0.50 & -0.001 (0.14) & 0.003 (0.17) & -0.001 (0.14)  & -0.001 (0.14) \\ 
0.8 & 0.78 & -0.002 (0.14) & 0.002 (0.19) & -0.001 (0.14) & -0.002 (0.14) \\ 
\hline \\[-1.8ex] 
\end{tabular} 
\end{table} 

\begin{table}[!htbp] \centering 
  \caption{Median bias (median absolute deviation) of the fixed effect $\theta_1$ in the bivariate model, $K=20$, average proportion of rejection 30\%. Full stands for the model with fully available data, Cca for complete case analysis, Mean for the sample average imputation of missing $\rho_w$, and Beta for the stochastic multiple imputation procedure for missing $\rho_w$.} 
  %\label{} 
\begin{tabular}{@{\extracolsep{5pt}} cccccc} 
\\[-1.8ex]\hline 
\hline \\[-1.8ex] 
$\rho_b$ & $\mathbb{E}[\rho_w]$ & Full & Cca & Mean & Beta \\ 
\hline \\[-1.8ex] 
-0.3 & 0 & 0.004 (0.17) & 0.007 (0.23) & 0.004 (0.17) &  0.004 (0.17) \\ 
-0.3 & 0.25 & -0.001 (0.14) & 0.001 (0.20) & -0.002 (0.14) &  -0.003 (0.14) \\ 
-0.3 & 0.50 & -0.000 (0.16) & 0.006 (0.24) & -0.001 (0.17) & -0.002 (0.16) \\ 
-0.3 & 0.78 & 0.001 (0.17) & 0.001 (0.24) & -0.000 (0.17)  & 0.001 (0.17) \\ 
0 & 0 & 0.004 (0.17) & 0.008 (0.24) & 0.004 (0.17) &  0.004 (0.17) \\ 
0 & 0.25 & -0.001 (0.17) & 0.001 (0.24) & -0.002 (0.17) & -0.003 (0.17) \\ 
0 & 0.50 & -0.002 (0.16) & 0.004 (0.24) & -0.001 (0.17) &  -0.002 (0.17) \\ 
0 & 0.78 & 0.001 (0.14) & -0.001 (0.19) & -0.001 (0.14) &  0.001 (0.14) \\ 
0.3 & 0 & 0.004 (0.14) & 0.010 (0.19) & 0.005 (0.14) &  0.005 (0.14) \\ 
0.3 & 0.25 & -0.002 (0.14) & 0.002 (0.19) & -0.000 (0.14) &  -0.002 (0.14) \\ 
0.3 & 0.50 & -0.003 (0.14) & -0.000 (0.19) & -0.001 (0.14) &  -0.001 (0.14) \\ 
0.3 & 0.78 & -0.002 (0.17) & -0.003 (0.24) & -0.003 (0.17) &  -0.000 (0.17) \\ 
0.5 & 0 & 0.005 (0.14) & 0.008 (0.20) & 0.005 (0.14) &  0.003 (0.14) \\ 
0.5 & 0.25 & -0.000 (0.17) & 0.001 (0.24) & -0.001 (0.17) &  0.001 (0.17) \\ 
0.5 & 0.50 & -0.003 (0.17) & -0.000 (0.24) & -0.001 (0.17) &  -0.001 (0.17) \\ 
0.5 & 0.78 & -0.002 (0.17) & -0.003 (0.24) & -0.003 (0.17)  & -0.001 (0.17) \\ 
0.8 & 0 & 0.004 (0.17) & 0.010 (0.24) & 0.005 (0.17) &  0.005 (0.17) \\ 
0.8 & 0.25 & 0.002 (0.17) & 0.002 (0.24) & 0.002 (0.17) & 0.001 (0.17) \\ 
0.8 & 0.50 & -0.002 (0.14) & 0.001 (0.20) & 0.001 (0.14) & -0.003 (0.14) \\ 
0.8 & 0.78 & -0.001 (0.14) & -0.004 (0.20) & -0.001 (0.14) &  0.000 (0.14) \\ 
\hline \\[-1.8ex] 
\end{tabular} 
\end{table}

\begin{table}[!htbp] \centering 
  \caption{Median bias (median absolute deviation) of the fixed effect $\theta_1$ in the bivariate model, $K=30$, average proportion of rejection 30\%. Full stands for the model with fully available data, Cca for complete case analysis, Mean for the sample average imputation of missing $\rho_w$, and Beta for the stochastic multiple imputation procedure for missing $\rho_w$.} 
  %\label{} 
\begin{tabular}{@{\extracolsep{5pt}} cccccc} 
\\[-1.8ex]\hline 
\hline \\[-1.8ex] 
$\rho_b$ & $\mathbb{E}[\rho_w]$ & Full & Cca & Mean & Beta \\ 
\hline \\[-1.8ex] 
-0.3 & 0 & 0.005 (0.14) & 0.002 (0.20) & 0.003 (0.14) &  0.003 (0.14) \\ 
-0.3 & 0.25 & 0.007 (0.14) & 0.007 (0.20) & 0.008 (0.14) & 0.007 (0.14) \\ 
-0.3 & 0.50 & 0.002 (0.17) & 0.005 (0.24) & 0.002 (0.17) & 0.002 (0.17) \\ 
-0.3 & 0.78 & -0.000 (0.14) & -0.002 (0.19) & 0.001 (0.14) &  -0.001 (0.14) \\ 
0 & 0 & 0.000 (0.17) & 0.001 (0.24) & 0.001 (0.17) &  0.002 (0.17) \\ 
0 & 0.25 & 0.006 (0.17) & 0.004 (0.24) & 0.004 (0.17) &  0.005 (0.17) \\ 
0 & 0.50 & -0.001 (0.17) & 0.005 (0.24) & -0.001 (0.17) &  -0.001 (0.17) \\ 
0 & 0.78 & -0.002 (0.17) & -0.003 (0.24) & -0.002 (0.17) &  -0.002 (0.17) \\ 
0.3 & 0 & 0.003 (0.17) & 0.001 (0.24) & 0.004 (0.17)  & 0.004 (0.17) \\ 
0.3 & 0.25 & 0.002 (0.14) & 0.007 (0.19) & 0.003 (0.14) &  0.004 (0.14) \\ 
0.3 & 0.50 & -0.002 (0.13) & 0.004 (0.20) & -0.003 (0.14) & -0.002 (0.14) \\ 
0.3 & 0.78 & -0.005 (0.13) & -0.004 (0.19) & -0.004 (0.14) &  -0.004 (0.13) \\ 
0.5 & 0 & 0.004 (0.14) & 0.002 (0.19) & 0.004 (0.13) &  0.005 (0.13) \\ 
0.5 & 0.25 & 0.003 (0.17) & 0.008 (0.24) & 0.004 (0.17)  & 0.005 (0.17) \\ 
0.5 & 0.50 & -0.001 (0.14) & 0.001 (0.20) & -0.003 (0.14) &  -0.001 (0.14) \\ 
0.5 & 0.78 & -0.004 (0.17) & -0.002 (0.20) & -0.004 (0.17) & -0.005 (0.17) \\ 
0.8 & 0 & 0.002 (0.17) & 0.003 (0.20) & 0.004 (0.17) &  0.004 (0.17) \\ 
0.8 & 0.25 & 0.003 (0.17) & 0.008 (0.20) & 0.005 (0.17)  & 0.004 (0.17) \\ 
0.8 & 0.50 & -0.001 (0.17) & 0.002 (0.20) & -0.002 (0.17) & -0.002 (0.17) \\ 
0.8 & 0.78 & -0.002 (0.17) & -0.003 (0.20) & -0.002 (0.17) & -0.003 (0.17) \\ 
\hline \\[-1.8ex] 
\end{tabular} 
\end{table}

\begin{table}[!htbp] \centering 
  \caption{Median bias (median absolute deviation) of the fixed effect $\theta_2$ in the bivariate model, $K=20$, average proportion of rejection 50\%. Full stands for the model with fully available data, Cca for complete case analysis, Mean for the sample average imputation of missing $\rho_w$, and Beta for the stochastic multiple imputation procedure for missing $\rho_w$.} 
  %\label{} 
\begin{tabular}{@{\extracolsep{5pt}} cccccc} 
\\[-1.8ex]\hline 
\hline \\[-1.8ex] 
$\rho_b$ & $\mathbb{E}[\rho_w]$ & Full & Cca & Mean & Beta \\ 
\hline \\[-1.8ex] 
-0.3 & 0 & 0.002 (0.11) & 0.008 (0.13) & 0.003 (0.11) &  0.001 (0.11) \\ 
-0.3 & 0.25 & 0.002 (0.11) & 0.004 (0.13) & 0.001 (0.11) &  0.002 (0.11) \\ 
-0.3 & 0.50 & 0.004 (0.10) & -0.004 (0.13) & 0.004 (0.10) &  0.004 (0.10) \\ 
-0.3 & 0.78 & 0.001 (0.11) & -0.002 (0.13) & 0.001 (0.11) &  0.001 (0.11) \\ 
0 & 0 & -0.001 (0.13) & 0.001 (0.19) & -0.001 (0.13) &  0.000 (0.13) \\ 
0 & 0.25 & -0.002 (0.11) & 0.001 (0.13) & -0.002 (0.11) & -0.003 (0.11) \\ 
0 & 0.50 & -0.002 (0.13) & -0.001 (0.16) & -0.001 (0.13) &  -0.000 (0.13) \\ 
0 & 0.78 & 0.001 (0.12) & -0.001 (0.15) & 0.003 (0.12) &  0.001 (0.12) \\ 
0.3 & 0 & 0.002 (0.12) & 0.001 (0.15) & 0.002 (0.12) &  0.000 (0.12) \\ 
0.3 & 0.25 & 0.000 (0.12) & 0.002 (0.15) & 0.001 (0.13) &  -0.000 (0.12) \\ 
0.3 & 0.50 & -0.002 (0.13) & 0.005 (0.15) & -0.002 (0.13) &  -0.003 (0.13) \\ 
0.3 & 0.78 & 0.001 (0.10) & 0.003 (0.12) & -0.002 (0.11) &  0.001 (0.10) \\ 
0.5 & 0 & -0.000 (0.11) & 0.002 (0.12) & 0.001 (0.11) &  -0.000 (0.11) \\ 
0.5 & 0.25 & 0.001 (0.10) & -0.000 (0.12) & 0.001 (0.11) &  0.000 (0.10) \\ 
0.5 & 0.50 & -0.001 (0.10) & 0.001 (0.12) & -0.002 (0.10) &  -0.001 (0.10) \\ 
0.5 & 0.78 & -0.002 (0.11) & 0.001 (0.15) & -0.003 (0.11) &  -0.002 (0.11) \\ 
0.8 & 0 & 0.002 (0.10) & 0.003 (0.12) & 0.003 (0.11) &  0.003 (0.10) \\ 
0.8 & 0.25 & 0.002 (0.13) & -0.004 (0.15) & 0.002 (0.13) &  0.002 (0.13) \\ 
0.8 & 0.50 & -0.001 (0.12) & -0.003 (0.15) & -0.000 (0.12) &  -0.001 (0.12) \\ 
0.8 & 0.78 & 0.000 (0.12) & -0.001 (0.15) & -0.002 (0.13) &  0.000 (0.13) \\ 
\hline \\[-1.8ex] 
\end{tabular} 
\end{table}

\begin{table}[!htbp] \centering 
  \caption{Median bias (median absolute deviation) of the fixed effect $\theta_2$ in the bivariate model, $K=30$, average proportion of rejection 50\%. Full stands for the model with fully available data, Cca for complete case analysis, Mean for the sample average imputation of missing $\rho_w$, and Beta for the stochastic multiple imputation procedure for missing $\rho_w$.} 
  %\label{} 
\begin{tabular}{@{\extracolsep{5pt}} cccccc} 
\\[-1.8ex]\hline 
\hline \\[-1.8ex] 
$\rho_b$ & $\mathbb{E}[\rho_w]$ & Full & Cca & Mean & Beta \\ 
\hline \\[-1.8ex] 
-0.3 & 0 & -0.001 (0.12) & -0.007 (0.15) & -0.001 (0.12) &  0.000 (0.12) \\ 
-0.3 & 0.25 & 0.003 (0.13) & -0.000 (0.16) & 0.003 (0.13) &  0.002 (0.13) \\ 
-0.3 & 0.50 & -0.002 (0.10) & -0.006 (0.13) & -0.002 (0.10) & -0.001 (0.10) \\ 
-0.3 & 0.78 & 0.003 (0.10) & 0.003 (0.12) & 0.001 (0.10) &  0.003 (0.10) \\ 
0 & 0 & 0.000 (0.10) & -0.001 (0.12) & 0.000 (0.10) &  0.000 (0.10) \\ 
0 & 0.25 & 0.004 (0.10) & 0.005 (0.13) & 0.004 (0.10) &  0.003 (0.10) \\ 
0 & 0.50 & -0.003 (0.11) & -0.003 (0.15) & -0.001 (0.11) &  -0.002 (0.11) \\ 
0 & 0.78 & 0.001 (0.10) & 0.004 (0.13) & 0.000 (0.10) &  0.000 (0.10) \\ 
0.3 & 0 & 0.002 (0.13) & 0.001 (0.15) & 0.002 (0.13) &  0.002 (0.13) \\ 
0.3 & 0.25 & 0.005 (0.12) & 0.006 (0.15) & 0.005 (0.12) &  0.005 (0.12) \\ 
0.3 & 0.50 & -0.001 (0.12) & -0.000 (0.15) & -0.003 (0.13) &  -0.001 (0.12) \\ 
0.3 & 0.78 & -0.001 (0.12) & 0.005 (0.15) & -0.002 (0.12) & -0.001 (0.12) \\ 
0.5 & 0 & 0.002 (0.13) & 0.002 (0.15) & 0.001 (0.13) &  0.002 (0.13) \\ 
0.5 & 0.25 & 0.005 (0.11) & 0.005 (0.13) & 0.004 (0.11) &  0.006 (0.11) \\ 
0.5 & 0.50 & -0.001 (0.11) & 0.003 (0.13) & -0.002 (0.11) & -0.002 (0.11) \\ 
0.5 & 0.78 & -0.000 (0.11) & 0.004 (0.13) & 0.000 (0.11) &  0.001 (0.11) \\ 
0.8 & 0 & 0.004 (0.11) & 0.004 (0.13) & 0.003 (0.11) &  0.004 (0.11) \\ 
0.8 & 0.25 & 0.003 (0.10) & 0.003 (0.15) & 0.003 (0.11) &  0.003 (0.10) \\ 
0.8 & 0.50 & -0.002 (0.11) & 0.002 (0.13) & -0.001 (0.11) &  -0.002 (0.11) \\ 
0.8 & 0.78 & 0.002 (0.11) & 0.003 (0.15) & -0.000 (0.11) &  0.001 (0.11) \\ 
\hline \\[-1.8ex] 
\end{tabular} 
\end{table}

\begin{table}[!htbp] \centering 
  \caption{Median bias (median absolute deviation) of the fixed effect $\theta_2$ in the bivariate model, $K=20$, average proportion of rejection 30\%. Full stands for the model with fully available data, Cca for complete case analysis, Mean for the sample average imputation of missing $\rho_w$, and Beta for the stochastic multiple imputation procedure for missing $\rho_w$.} 
  %\label{} 
\begin{tabular}{@{\extracolsep{5pt}} cccccc} 
\\[-1.8ex]\hline 
\hline \\[-1.8ex] 
$\rho_b$ & $\mathbb{E}[\rho_w]$ & Full & Cca & Mean & Beta \\ 
\hline \\[-1.8ex] 
-0.3 & 0 & 0.001 (0.13) & -0.000 (0.19) & 0.002 (0.13) &  0.002 (0.13) \\ 
-0.3 & 0.25 & 0.002 (0.11) & -0.004 (0.15) & 0.001 (0.11) &  0.001 (0.11) \\ 
-0.3 & 0.50 & 0.004 (0.13) & 0.001 (0.18) & 0.005 (0.13) &  0.006 (0.13) \\ 
-0.3 & 0.78 & 0.001 (0.12) & -0.000 (0.18) & 0.001 (0.12)  & 0.001 (0.12) \\ 
0 & 0 & -0.000 (0.12) & -0.002 (0.18) & 0.001 (0.12) &  0.001 (0.12) \\ 
0 & 0.25 & -0.002 (0.12) & -0.006 (0.18) & -0.002 (0.13) &  -0.001 (0.12) \\ 
0 & 0.50 & -0.001 (0.13) & 0.001 (0.19) & -0.001 (0.13) &  -0.001 (0.13) \\ 
0 & 0.78 & 0.001 (0.10) & 0.002 (0.15) & 0.001 (0.11) &  0.002 (0.10) \\ 
0.3 & 0 & 0.002 (0.11) & -0.001 (0.15) & 0.002 (0.11) &  0.002 (0.11) \\ 
0.3 & 0.25 & 0.000 (0.10) & -0.006 (0.15) & 0.000 (0.10) &  -0.001 (0.11) \\ 
0.3 & 0.50 & -0.001 (0.10) & 0.000 (0.15) & -0.002 (0.10)  & -0.001 (0.11) \\ 
0.3 & 0.78 & 0.001 (0.13) & 0.003 (0.18) & 0.001 (0.13) &  0.001 (0.13) \\ 
0.5 & 0 & 0.000 (0.10) & -0.002 (0.15) & 0.000 (0.10) &  0.001 (0.10) \\ 
0.5 & 0.25 & 0.000 (0.13) & -0.006 (0.19) & 0.001 (0.13) &  -0.000 (0.13) \\ 
0.5 & 0.50 & -0.001 (0.12) & 0.002 (0.19) & -0.003 (0.12)  & -0.001 (0.13) \\ 
0.5 & 0.78 & -0.002 (0.12) & 0.003 (0.18) & -0.001 (0.13) &  -0.002 (0.13) \\ 
0.8 & 0 & 0.002 (0.12) & 0.000 (0.19) & 0.002 (0.12) &  0.002 (0.12) \\ 
0.8 & 0.25 & 0.002 (0.13) & -0.001 (0.19) & 0.000 (0.13)  & 0.003 (0.13) \\ 
0.8 & 0.50 & -0.001 (0.10) & -0.001 (0.15) & -0.003 (0.10) &  0.001 (0.10) \\ 
0.8 & 0.78 & 0.001 (0.10) & 0.003 (0.15) & 0.001 (0.11) &  0.001 (0.10) \\ 
\hline \\[-1.8ex] 
\end{tabular} 
\end{table} 

\begin{table}[!htbp] \centering 
  \caption{Median bias (median absolute deviation) of the fixed effect $\theta_2$ in the bivariate model, $K=30$, average proportion of rejection 30\%. Full stands for the model with fully available data, Cca for complete case analysis, Mean for the sample average imputation of missing $\rho_w$, and Beta for the stochastic multiple imputation procedure for missing $\rho_w$.} 
  %\label{} 
\begin{tabular}{@{\extracolsep{5pt}} cccccc} 
\\[-1.8ex]\hline 
\hline \\[-1.8ex] 
$\rho_b$ & $\mathbb{E}[\rho_w]$ & Full & Cca & Mean & Beta \\  
\hline \\[-1.8ex] 
-0.3 & 0 & -0.001 (0.10) & -0.003 (0.15) & -0.001 (0.10)  & -0.001 (0.10) \\ 
-0.3 & 0.25 & 0.003 (0.10) & -0.001 (0.15) & 0.004 (0.11) &  0.002 (0.11) \\ 
-0.3 & 0.50 & -0.002 (0.13) & -0.007 (0.18) & -0.003 (0.13) &  -0.001 (0.13) \\ 
-0.3 & 0.78 & 0.003 (0.10) & 0.002 (0.15) & 0.003 (0.10) &  0.003 (0.10) \\ 
0 & 0 & 0.000 (0.13) & -0.003 (0.18) & 0.001 (0.13) &  0.001 (0.13) \\ 
0 & 0.25 & 0.004 (0.12) & 0.001 (0.18) & 0.005 (0.12) &  0.004 (0.12) \\ 
0 & 0.50 & -0.003 (0.12) & -0.004 (0.18) & -0.003 (0.12) &  -0.003 (0.12) \\ 
0 & 0.78 & 0.001 (0.12) & -0.001 (0.18) & -0.000 (0.12) &  0.001 (0.12) \\ 
0.3 & 0 & 0.002 (0.13) & 0.001 (0.18) & 0.003 (0.13) &  0.003 (0.13) \\ 
0.3 & 0.25 & 0.005 (0.11) & 0.002 (0.15) & 0.005 (0.11) &  0.005 (0.11) \\ 
0.3 & 0.50 & -0.001 (0.11) & 0.003 (0.15) & -0.001 (0.11) &  -0.001 (0.11) \\ 
0.3 & 0.78 & -0.001 (0.11) & -0.002 (0.15) & -0.001 (0.11) & -0.001 (0.11) \\ 
0.5 & 0 & 0.002 (0.11) & 0.001 (0.15) & 0.002 (0.11) &  0.002 (0.11) \\ 
0.5 & 0.25 & 0.005 (0.13) & 0.003 (0.18) & 0.005 (0.13)  & 0.005 (0.13) \\ 
0.5 & 0.50 & -0.001 (0.11) & 0.004 (0.15) & -0.003 (0.11)  & -0.001 (0.11) \\ 
0.5 & 0.78 & -0.000 (0.13) & -0.001 (0.15) & 0.002 (0.13) &  0.000 (0.13) \\ 
0.8 & 0 & 0.004 (0.13) & 0.006 (0.15) & 0.005 (0.13) &  0.005 (0.13) \\ 
0.8 & 0.25 & 0.003 (0.13) & 0.003 (0.15) & 0.003 (0.13) &  0.004 (0.13) \\ 
0.8 & 0.50 & -0.002 (0.13) & 0.005 (0.15) & -0.002 (0.13) &  -0.003 (0.13) \\ 
0.8 & 0.78 & 0.002 (0.13) & -0.002 (0.15) & 0.002 (0.13) & 0.002 (0.13) \\ 
\hline \\[-1.8ex] 
\end{tabular} 
\end{table}

\begin{table}[!htbp] \centering 
  \caption{Median distance (median absolute deviation) of the true sampling correlation coefficient between the estimators of the main effects $(\theta_1, \theta_2)$ in the bivariate model, $K=20$, average proportion of rejection 50\%. Full stands for the model with fully available data, Cca for complete case analysis, Mean for the sample average imputation of missing $\rho_w$, and Beta for the stochastic multiple imputation procedure for missing $\rho_w$.} 
  %\label{} 
\begin{tabular}{@{\extracolsep{5pt}} cccccc} 
\\[-1.8ex]\hline 
\hline \\[-1.8ex] 
$\rho_b$ & $\mathbb{E}[\rho_w]$ & Full & Cca & Mean & Beta \\ 
\hline \\[-1.8ex] 
-0.3 & 0 & 0 (0.13) & -0.030 (0.15) & 0.003 (0.13) &  0.002 (0.13) \\ 
-0.3 & 0.25 & 0 (0.09) & -0.029 (0.11) & 0.002 (0.09) &  0.000 (0.09) \\ 
-0.3 & 0.50 & 0 (0.05) & -0.027 (0.06) & 0.002 (0.05) & 0.002 (0.05) \\ 
-0.3 & 0.78 & 0 (0.11) & -0.012 (0.13) & 0.011 (0.11) &  0.003 (0.11) \\ 
0 & 0 & 0 (0.16) & -0.013 (0.25) & 0.001 (0.17) &  -0.002 (0.16) \\ 
0 & 0.25 & 0 (0.13) & -0.016 (0.16) & -0.002 (0.13) &  -0.002 (0.13) \\ 
0 & 0.50 & 0 (0.16) & -0.006 (0.20) & -0.002 (0.16) &  -0.003 (0.16) \\ 
0 & 0.78 & 0 (0.12) & -0.005 (0.15) & -0.001 (0.13) &  -0.001 (0.12) \\ 
0.3 & 0 & 0 (0.08) & -0.005 (0.10) & -0.002 (0.08) &  -0.002 (0.08) \\ 
0.3 & 0.25 & 0 (0.15) & 0.009 (0.18) & -0.002 (0.15) &  -0.000 (0.15) \\ 
0.3 & 0.50 & 0 (0.16) & 0.016 (0.21) & -0.000 (0.16) &  -0.000 (0.16) \\ 
0.3 & 0.78 & 0 (0.13) & 0.003 (0.16) & -0.004 (0.13) &  -0.004 (0.13) \\ 
0.5 & 0 & 0 (0.10) & -0.003 (0.12) & -0.004 (0.10) &  -0.002 (0.10) \\ 
0.5 & 0.25 & 0 (0.06) & 0.008 (0.08) & -0.003 (0.07) &  -0.001 (0.06) \\ 
0.5 & 0.50 & 0 (0.11) & 0.017 (0.14) & -0.004 (0.12) &  -0.005 (0.11) \\ 
0.5 & 0.78 & 0 (0.13) & 0.011 (0.19) & -0.008 (0.13) &  -0.006 (0.13) \\ 
0.8 & 0 & 0 (0.13) & -0.020 (0.16) & -0.008 (0.13) & -0.006 (0.13) \\ 
0.8 & 0.25 & 0 (0.17) & -0.002 (0.21) & -0.009 (0.16) &  -0.008 (0.17) \\ 
0.8 & 0.50 & 0 (0.13) & 0.011 (0.16) & -0.010 (0.13) &  -0.008 (0.13) \\ 
0.8 & 0.78 & 0 (0.09) & 0.017 (0.11) & -0.013 (0.09) &  -0.008 (0.09) \\ 
\hline \\[-1.8ex] 
\end{tabular} 
\end{table}

\begin{table}[!htbp] \centering 
  \caption{Median distance (median absolute deviation) of the true sampling correlation coefficient between the estimators of the main effects $(\theta_1, \theta_2)$ in the bivariate model, $K=30$, average proportion of rejection 50\%. Full stands for the model with fully available data, Cca for complete case analysis, Mean for the sample average imputation of missing $\rho_w$, and Beta for the stochastic multiple imputation procedure for missing $\rho_w$.} 
  %\label{} 
\begin{tabular}{@{\extracolsep{5pt}} cccccc} 
\\[-1.8ex]\hline 
\hline \\[-1.8ex] 
$\rho_b$ & $\mathbb{E}[\rho_w]$ & Full & Cca & Mean & Beta \\ 
\hline \\[-1.8ex] 
-0.3 & 0 & 0 (0.15) & -0.020 (0.19) & 0.002 (0.15) &  0.000 (0.15) \\ 
-0.3 & 0.25 & 0 (0.16) & -0.017 (0.21) & 0.000 (0.17) &  -0.001 (0.17) \\ 
-0.3 & 0.50 & 0 (0.13) & -0.015 (0.16) & 0.003 (0.13) &  0.001 (0.13) \\ 
-0.3 & 0.78 & 0 (0.11) & -0.003 (0.13) & 0.004 (0.11) &  0.001 (0.11) \\ 
0 & 0 & 0 (0.07) & -0.010 (0.09) & 0.002 (0.08) &  0.002 (0.07) \\ 
0 & 0.25 & 0 (0.12) & -0.010 (0.15) & -0.003 (0.12)  & 0.000 (0.12) \\ 
0 & 0.50 & 0 (0.09) & -0.008 (0.14) & 0.000 (0.09) & -0.002 (0.09) \\ 
0 & 0.78 & 0 (0.13) & -0.005 (0.16) & -0.003 (0.13) &  -0.001 (0.13) \\ 
0.3 & 0 & 0 (0.16) & -0.004 (0.20) & -0.000 (0.16) &  -0.001 (0.16) \\ 
0.3 & 0.25 & 0 (0.14) & -0.002 (0.17) & -0.005 (0.14)  & -0.002 (0.14) \\ 
0.3 & 0.50 & 0 (0.10) & -0.004 (0.12) & -0.004 (0.10) &  -0.003 (0.10) \\ 
0.3 & 0.78 & 0 (0.15) & -0.005 (0.19) & -0.009 (0.15) & -0.001 (0.15) \\ 
0.5 & 0 & 0 (0.15) & -0.003 (0.20) & -0.002 (0.15) &  -0.003 (0.15) \\ 
0.5 & 0.25 & 0 (0.13) & -0.001 (0.16) & -0.004 (0.13) &  -0.002 (0.13) \\ 
0.5 & 0.50 & 0 (0.12) & 0.001 (0.14) & -0.005 (0.12) &  -0.002 (0.11) \\ 
0.5 & 0.78 & 0 (0.08) & 0.006 (0.10) & -0.012 (0.08) &  -0.004 (0.08) \\ 
0.8 & 0 & 0 (0.13) & -0.010 (0.15) & -0.005 (0.13) &  -0.005 (0.12) \\ 
0.8 & 0.25 & 0 (0.05) & -0.003 (0.08) & -0.007 (0.06) &  -0.006 (0.05) \\ 
0.8 & 0.50 & 0 (0.12) & 0.004 (0.15) & -0.009 (0.12) &  -0.008 (0.12) \\ 
0.8 & 0.78 & 0 (0.11) & 0.005 (0.17) & -0.019 (0.11) &  -0.010 (0.11) \\ 
\hline \\[-1.8ex] 
\end{tabular} 
\end{table}

\begin{table}[!htbp] \centering 
  \caption{Median distance (median absolute deviation) of the true sampling correlation coefficient between the estimators of the main effects $(\theta_1, \theta_2)$ in the bivariate model, $K=20$, average proportion of rejection 30\%. Full stands for the model with fully available data, Cca for complete case analysis, Mean for the sample average imputation of missing $\rho_w$, and Beta for the stochastic multiple imputation procedure for missing $\rho_w$.} 
  %\label{} 
\begin{tabular}{@{\extracolsep{5pt}} cccccc} 
\\[-1.8ex]\hline 
\hline \\[-1.8ex] 
$\rho_b$ & $\mathbb{E}[\rho_w]$ & Full & Cca & Mean & Beta \\ 
\hline \\[-1.8ex] 
-0.3 & 0 & 0 (0.16) & -0.017 (0.24) & 0.001 (0.16) & 0.001 (0.16) \\ 
-0.3 & 0.25 & 0 (0.13) & -0.012 (0.20) & 0.000 (0.13)  & 0.002 (0.13) \\ 
-0.3 & 0.50 & 0 (0.16) & -0.009 (0.25) & 0.002 (0.16) &  -0.000 (0.16) \\ 
-0.3 & 0.78 & 0 (0.12) & -0.005 (0.19) & 0.003 (0.12) &  0.002 (0.12) \\ 
0 & 0 & 0 (0.08) & -0.010 (0.11) & 0.000 (0.08) &  0.000 (0.08) \\ 
0 & 0.25 & 0 (0.15) & -0.010 (0.22) & -0.003 (0.14) &  -0.002 (0.15) \\ 
0 & 0.50 & 0 (0.16) & -0.006 (0.25) & -0.002 (0.16) &  -0.002 (0.16) \\ 
0 & 0.78 & 0 (0.13) & -0.002 (0.19) & -0.002 (0.13) &  -0.002 (0.13) \\ 
0.3 & 0 & 0 (0.10) & -0.003 (0.15) & -0.002 (0.10) & -0.002 (0.10) \\ 
0.3 & 0.25 & 0 (0.06) & -0.000 (0.10) & 0.000 (0.07) &  -0.000 (0.06) \\ 
0.3 & 0.50 & 0 (0.11) & -0.005 (0.18) & -0.001 (0.11) &  0.000 (0.11) \\ 
0.3 & 0.78 & 0 (0.11) & 0.005 (0.17) & -0.002 (0.11) & -0.000 (0.11) \\ 
0.5 & 0 & 0 (0.13) & -0.004 (0.19) & -0.002 (0.13) & -0.001 (0.13) \\ 
0.5 & 0.25 & 0 (0.17) & 0.002 (0.26) & -0.002 (0.16) &  -0.000 (0.17) \\ 
0.5 & 0.50 & 0 (0.13) & -0.000 (0.20) & -0.004 (0.13) &  -0.001 (0.13) \\ 
0.5 & 0.78 & 0 (0.09) & 0.005 (0.13) & -0.006 (0.09) &  -0.003 (0.09) \\ 
0.8 & 0 & 0 (0.15) & -0.008 (0.23) & -0.004 (0.15) & -0.003 (0.15) \\ 
0.8 & 0.25 & 0 (0.16) & -0.004 (0.25) & -0.005 (0.16)  & -0.003 (0.16) \\ 
0.8 & 0.50 & 0 (0.13) & 0.001 (0.20) & -0.005 (0.13) &  -0.004 (0.13) \\ 
0.8 & 0.78 & 0 (0.11) & 0.009 (0.16) & -0.007 (0.11) &  -0.004 (0.11) \\ 
\hline \\[-1.8ex] 
\end{tabular} 
\end{table}

\begin{table}[!htbp] \centering 
  \caption{Median distance (median absolute deviation) of the true sampling correlation coefficient between the estimators of the main effects $(\theta_1, \theta_2)$ in the bivariate model, $K=30$, average proportion of rejection 30\%. Full stands for the model with fully available data, Cca for complete case analysis, Mean for the sample average imputation of missing $\rho_w$, and Beta for the stochastic multiple imputation procedure for missing $\rho_w$.} 
  %\label{} 
\begin{tabular}{@{\extracolsep{5pt}} cccccc} 
\\[-1.8ex]\hline 
\hline \\[-1.8ex] 
$\rho_b$ & $\mathbb{E}[\rho_w]$ & Full & Cca & Mean & Beta \\ 
\hline \\[-1.8ex] 
-0.3 & 0 & 0 (0.07) & -0.013 (0.11) & 0.001 (0.07) &  0.001 (0.07) \\ 
-0.3 & 0.25 & 0 (0.12) & -0.015 (0.18) & 0.001 (0.12) & -0.000 (0.12) \\ 
-0.3 & 0.50 & 0 (0.06) & -0.009 (0.09) & 0.001 (0.07) & 0.001 (0.06) \\ 
-0.3 & 0.78 & 0 (0.13) & -0.004 (0.19) & 0.005 (0.13) &  0.002 (0.13) \\ 
0 & 0 & 0 (0.16) & -0.002 (0.25) & 0.001 (0.16) &  0.002 (0.17) \\ 
0 & 0.25 & 0 (0.14) & -0.014 (0.21) & -0.002 (0.14) & -0.002 (0.14) \\ 
0 & 0.50 & 0 (0.10) & -0.010 (0.15) & -0.001 (0.10) &  -0.001 (0.10) \\ 
0 & 0.78 & 0 (0.15) & 0.006 (0.24) & -0.001 (0.15) &  -0.001 (0.15) \\ 
0.3 & 0 & 0 (0.15) & -0.005 (0.24) & -0.001 (0.15) &  -0.001 (0.15) \\ 
0.3 & 0.25 & 0 (0.13) & -0.002 (0.20) & -0.002 (0.13) &  -0.001 (0.13) \\ 
0.3 & 0.50 & 0 (0.12) & -0.003 (0.17) & -0.002 (0.11) &  -0.001 (0.11) \\ 
0.3 & 0.78 & 0 (0.08) & 0.003 (0.12) & -0.003 (0.08) &  -0.001 (0.08) \\ 
0.5 & 0 & 0 (0.13) & -0.001 (0.20) & 0.000 (0.13) &  -0.000 (0.13) \\ 
0.5 & 0.25 & 0 (0.13) & -0.002 (0.20) & -0.003 (0.14) &  -0.002 (0.14) \\ 
0.5 & 0.50 & 0 (0.12) & -0.000 (0.19) & -0.004 (0.12) &  -0.002 (0.12) \\ 
0.5 & 0.78 & 0 (0.16) & 0.003 (0.20) & -0.006 (0.16) &  -0.003 (0.16) \\ 
0.8 & 0 & 0 (0.11) & -0.006 (0.14) & -0.002 (0.12) &  -0.002 (0.11) \\ 
0.8 & 0.25 & 0 (0.06) & -0.001 (0.08) & -0.005 (0.07) &  -0.003 (0.06) \\ 
0.8 & 0.50 & 0 (0.13) & -0.001 (0.17) & -0.008 (0.14) &  -0.004 (0.14) \\ 
0.8 & 0.78 & 0 (0.16) & 0.002 (0.20) & -0.011 (0.16) &  -0.005 (0.16) \\ 
\hline \\[-1.8ex] 
\end{tabular} 
\end{table} 

\newpage

\subsection{Surrogate model analysis}

\begin{table}[!htbp] \centering 
	\caption{Median bias (median absolute deviation) of the regression coefficient $\delta_1$ in the surrogate model, $K=20$, average proportion of rejection 50\%.  Full stands for the model with fully available data, Cca for complete case analysis, Mean for the sample average imputation of missing $\rho_w$, and Beta for the stochastic multiple imputation procedure for missing $\rho_w$.} 
	\label{tab:surrog_beta1} 
	\begin{tabular}{@{\extracolsep{5pt}} cccccc} 
		\\[-1.8ex]\hline 
		\hline \\[-1.8ex] 
		$\rho_b$ & $\mathbb{E}[\rho_w]$ & Full & Cca & Mean & Beta \\ 
		\hline \\[-1.8ex] 
		-0.4 & 0 & -0.008 (0.25) & -0.036 (0.31) & 0.007 (0.24) &  -0.004 (0.25) \\ 
		-0.4 & 0.25 & -0.023 (0.21) & -0.068 (0.26) & 0.043 (0.21) &  -0.013 (0.21) \\ 
		-0.4 & 0.50 & -0.064 (0.17) & -0.107 (0.21) & 0.043 (0.17) &  -0.056 (0.17) \\ 
		-0.4 & 0.78 & -0.055 (0.12) & -0.129 (0.15) & 0.102 (0.12) &  -0.046 (0.12) \\ 
		0 & 0 & 0.001 (0.33) & -0.007 (0.50) & 0.012 (0.31) &  -0.001 (0.33) \\ 
		0 & 0.25 & -0.017 (0.26) & -0.036 (0.32) & 0.049 (0.25) & -0.012 (0.26) \\ 
		0 & 0.50 & -0.032 (0.32) & -0.055 (0.40) & 0.077 (0.31) &  -0.029 (0.32) \\ 
		0 & 0.78 & -0.029 (0.28) & -0.063 (0.37) & 0.118 (0.28) &  -0.022 (0.28) \\ 
		0.4 & 0 & 0.006 (0.23) & 0.028 (0.30) & 0.021 (0.24) &  0.007 (0.23) \\ 
		0.4 & 0.25 & 0.002 (0.18) & 0.034 (0.23) & 0.057 (0.19) &  -0.002 (0.18) \\ 
		0.4 & 0.50 & 0.007 (0.32) & 0.029 (0.41) & 0.105 (0.31) &  0.007 (0.31) \\ 
		0.4 & 0.78 & -0.009 (0.24) & -0.003 (0.30) & 0.114 (0.24) &  -0.007 (0.24) \\ 
		0.7 & 0 & 0.022 (0.21) & 0.064 (0.27) & 0.040 (0.21) &  0.019 (0.21) \\ 
		0.7 & 0.25 & 0.020 (0.18) & 0.070 (0.23) & 0.071 (0.18) &  0.014 (0.18) \\ 
		0.7 & 0.50 & 0.026 (0.14) & 0.067 (0.18) & 0.109 (0.15) &  0.020 (0.14) \\ 
		0.7 & 0.78 & 0.013 (0.25) & 0.044 (0.38) & 0.110 (0.24) &  0.006 (0.25) \\ 
		0.9 & 0 & 0.028 (0.24) & 0.025 (0.30) & 0.036 (0.24)  & 0.021 (0.24) \\ 
		0.9 & 0.25 & 0.019 (0.30) & 0.044 (0.39) & 0.049 (0.30)  & 0.012 (0.30) \\ 
		0.9 & 0.50 & 0.027 (0.29) & 0.055 (0.37) & 0.070 (0.29)  & 0.015 (0.28) \\ 
		0.9 & 0.78 & 0.013 (0.26) & 0.033 (0.33) & 0.049 (0.26) & 0.009 (0.25) \\ 
		\hline \\[-1.8ex] 
	\end{tabular} 
\end{table}

\begin{table}[!htbp] \centering 
	\caption{Median bias (median absolute deviation) of the regression coefficient $\delta_1$ in the surrogate model, $K=30$, average proportion of rejection 50\%.  Full stands for the model with fully available data, Cca for complete case analysis, Mean for the sample average imputation of missing $\rho_w$, and Beta for the stochastic multiple imputation procedure for missing $\rho_w$.} 
	\label{tab:surrog_beta2} 
	\begin{tabular}{@{\extracolsep{5pt}} cccccc} 
		\\[-1.8ex]\hline 
		\hline \\[-1.8ex] 
		$\rho_b$ & $\mathbb{E}[\rho_w]$ & Full & Cca & Mean & Beta \\ 
		\hline \\[-1.8ex] 
		-0.4 & 0 & -0.026 (0.20) & -0.033 (0.26) & -0.012 (0.20) &  -0.021 (0.20) \\ 
		-0.4 & 0.25 & -0.024 (0.30) & -0.042 (0.39) & 0.036 (0.30) &  -0.019 (0.30) \\ 
		-0.4 & 0.50 & -0.024 (0.24) & -0.057 (0.30) & 0.084 (0.24) &  -0.021 (0.24) \\ 
		-0.4 & 0.78 & -0.040 (0.22) & -0.092 (0.28) & 0.119 (0.22) &  -0.029 (0.22) \\ 
		0 & 0 & -0.003 (0.20) & 0.012 (0.24) & 0.010 (0.20) &  -0.002 (0.20) \\ 
		0 & 0.25 & -0.001 (0.16) & 0.002 (0.20) & 0.061 (0.16) &  0.001 (0.16) \\ 
		0 & 0.50 & -0.021 (0.21) & -0.032 (0.34) & 0.084 (0.22) &  -0.018 (0.21) \\ 
		0 & 0.78 & -0.029 (0.24) & -0.053 (0.29) & 0.122 (0.24) & -0.019 (0.24) \\ 
		0.4 & 0 & 0.028 (0.29) & 0.034 (0.37) & 0.038 (0.30) &  0.022 (0.29) \\ 
		0.4 & 0.25 & 0.013 (0.29) & 0.022 (0.37) & 0.068 (0.29) &  0.011 (0.29) \\ 
		0.4 & 0.50 & -0.002 (0.27) & 0.003 (0.33) & 0.093 (0.27) & -0.003 (0.27) \\ 
		0.4 & 0.78 & -0.003 (0.21) & -0.005 (0.27) & 0.126 (0.22) & -0.004 (0.22) \\ 
		0.7 & 0 & 0.037 (0.29) & 0.060 (0.37) & 0.046 (0.29) &  0.026 (0.29) \\ 
		0.7 & 0.25 & 0.024 (0.25) & 0.045 (0.30) & 0.075 (0.25) &  0.016 (0.25) \\ 
		0.7 & 0.50 & 0.020 (0.24) & 0.034 (0.30) & 0.094 (0.24) &  0.014 (0.24) \\ 
		0.7 & 0.78 & 0.008 (0.21) & 0.023 (0.26) & 0.101 (0.21) &  0.005 (0.20) \\ 
		0.9 & 0 & 0.027 (0.18) & 0.028 (0.21) & 0.032 (0.17) &  0.017 (0.17) \\ 
		0.9 & 0.25 & 0.024 (0.17) & 0.038 (0.26) & 0.054 (0.18)  & 0.012 (0.17) \\ 
		0.9 & 0.50 & 0.022 (0.23) & 0.041 (0.29) & 0.069 (0.23) &  0.012 (0.23) \\ 
		0.9 & 0.78 & 0.013 (0.12) & 0.033 (0.19) & 0.052 (0.13) &  0.001 (0.12) \\ 
		\hline \\[-1.8ex] 
	\end{tabular} 
\end{table}
 
 \begin{table}[!htbp] \centering 
 	\caption{Median bias (median absolute deviation) of the regression coefficient $\delta_1$ in the surrogate model, $K=20$, average proportion of rejection 30\%.  Full stands for the model with fully available data, Cca for complete case analysis, Mean for the sample average imputation of missing $\rho_w$, and Beta for the stochastic multiple imputation procedure for missing $\rho_w$.}
 	\label{tab:surrog_beta3} 
 	\begin{tabular}{@{\extracolsep{5pt}} cccccc} 
 		\\[-1.8ex]\hline 
 		\hline \\[-1.8ex] 
 	$\rho_b$ & $\mathbb{E}[\rho_w]$ & Full & Cca & Mean & Beta \\
 		\hline \\[-1.8ex] 
 		-0.4 & 0 & -0.008 (0.32) & -0.014 (0.50) & 0.006 (0.30) &  -0.005 (0.32) \\ 
 		-0.4 & 0.25 & -0.022 (0.26) & -0.035 (0.39) & 0.020 (0.25) & -0.014 (0.26) \\ 
 		-0.4 & 0.50 & -0.063 (0.32) & -0.084 (0.52) & 0.007 (0.32) &  -0.058 (0.32) \\ 
 		-0.4 & 0.78 & -0.055 (0.28) & -0.080 (0.46) & 0.043 (0.29) &  -0.051 (0.28) \\ 
 		0 & 0 & 0.002 (0.23) & -0.007 (0.39) & 0.013 (0.24) &  0.001 (0.24) \\ 
 		0 & 0.25 & -0.014 (0.19) & -0.013 (0.28) & 0.024 (0.18)  & -0.010 (0.19) \\ 
 		0 & 0.50 & -0.032 (0.32) & -0.023 (0.50) & 0.040 (0.31) &  -0.027 (0.32) \\ 
 		0 & 0.78 & -0.029 (0.24) & -0.048 (0.37) & 0.061 (0.24) &  -0.025 (0.24) \\ 
 		0.4 & 0 & 0.006 (0.21) & 0.020 (0.35) & 0.019 (0.22) &  0.005 (0.22) \\ 
 		0.4 & 0.25 & 0.002 (0.18) & 0.015 (0.29) & 0.038 (0.18) &  0.004 (0.18) \\ 
 		0.4 & 0.50 & 0.007 (0.14) & 0.011 (0.22) & 0.072 (0.15) &  0.008 (0.14) \\ 
 		0.4 & 0.78 & -0.013 (0.27) & -0.020 (0.44) & 0.068 (0.27) &  -0.011 (0.27) \\ 
 		0.7 & 0 & 0.024 (0.24) & 0.044 (0.37) & 0.037 (0.23)  & 0.021 (0.24) \\ 
 		0.7 & 0.25 & 0.020 (0.30) & 0.037 (0.50) & 0.052 (0.30) &  0.017 (0.30) \\ 
 		0.7 & 0.50 & 0.027 (0.29) & 0.040 (0.48) & 0.079 (0.29) &  0.027 (0.29) \\ 
 		0.7 & 0.78 & 0.013 (0.26) & 0.013 (0.42) & 0.065 (0.26) &  0.009 (0.26) \\ 
 		0.9 & 0 & 0.027 (0.20) & 0.032 (0.31) & 0.033 (0.20) &  0.023 (0.21) \\ 
 		0.9 & 0.25 & 0.018 (0.30) & 0.034 (0.48) & 0.033 (0.30)  & 0.012 (0.30) \\ 
 		0.9 & 0.50 & 0.029 (0.24) & 0.040 (0.38) & 0.055 (0.24)  & 0.022 (0.24) \\ 
 		0.9 & 0.78 & 0.013 (0.22) & 0.026 (0.35) & 0.035 (0.23)  & 0.012 (0.22) \\ 
 		\hline \\[-1.8ex] 
 	\end{tabular} 
 \end{table}

 \begin{table}[!htbp] \centering 
 	\caption{Median bias (median absolute deviation) of the regression coefficient $\delta_1$ in the surrogate model, $K=30$, average proportion of rejection 30\%.  Full stands for the model with fully available data, Cca for complete case analysis, Mean for the sample average imputation of missing $\rho_w$, and Beta for the stochastic multiple imputation procedure for missing $\rho_w$.} 
 	\label{tab:surrog_beta4} 
 	\begin{tabular}{@{\extracolsep{5pt}} cccccc} 
 		\\[-1.8ex]\hline 
 		\hline \\[-1.8ex] 
 		$\rho_b$ & $\mathbb{E}[\rho_w]$ & Full & Cca & Mean  & Beta \\
 		\hline \\[-1.8ex] 
 		-0.4 & 0 & -0.026 (0.20) & -0.043 (0.30) & -0.014 (0.20) &  -0.024 (0.20) \\ 
 		-0.4 & 0.25 & -0.024 (0.16) & -0.041 (0.24) & 0.021 (0.16) &  -0.019 (0.16) \\ 
 		-0.4 & 0.50 & -0.024 (0.22) & -0.045 (0.35) & 0.044 (0.23) &  -0.022 (0.22) \\ 
 		-0.4 & 0.78 & -0.040 (0.24) & -0.064 (0.36) & 0.062 (0.24) &  -0.033 (0.24) \\ 
 		0 & 0 & -0.003 (0.29) & 0.001 (0.48) & 0.009 (0.30)  & -0.004 (0.29) \\ 
 		0 & 0.25 & -0.001 (0.29) & -0.020 (0.48) & 0.038 (0.30) &  0.000 (0.29) \\ 
 		0 & 0.50 & -0.021 (0.27) & -0.032 (0.43) & 0.047 (0.27) &  -0.023 (0.27) \\ 
 		0 & 0.78 & -0.029 (0.21) & -0.030 (0.33) & 0.063 (0.22) &  -0.030 (0.22) \\ 
 		0.4 & 0 & 0.028 (0.29) & 0.039 (0.48) & 0.037 (0.29) &  0.023 (0.29) \\ 
 		0.4 & 0.25 & 0.013 (0.25) & 0.023 (0.38) & 0.053 (0.25) &  0.011 (0.25) \\ 
 		0.4 & 0.50 & -0.002 (0.24) & 0.003 (0.36) & 0.059 (0.24)  & -0.003 (0.24) \\ 
 		0.4 & 0.78 & -0.003 (0.21) & -0.006 (0.32) & 0.072 (0.21)  & -0.001 (0.21) \\ 
 		0.7 & 0 & 0.037 (0.18) & 0.047 (0.26) & 0.048 (0.17)  & 0.035 (0.17) \\ 
 		0.7 & 0.25 & 0.024 (0.15) & 0.042 (0.24) & 0.055 (0.16) &  0.021 (0.16) \\ 
 		0.7 & 0.50 & 0.020 (0.23) & 0.029 (0.35) & 0.067 (0.23) &  0.013 (0.23) \\ 
 		0.7 & 0.78 & 0.008 (0.32) & 0.018 (0.40) & 0.061 (0.30) &  0.008 (0.31) \\ 
 		0.9 & 0 & 0.027 (0.27) & 0.024 (0.35) & 0.030 (0.27) &  0.021 (0.27) \\ 
 		0.9 & 0.25 & 0.024 (0.22) & 0.033 (0.28) & 0.043 (0.22) & 0.019 (0.22) \\ 
 		0.9 & 0.50 & 0.022 (0.15) & 0.034 (0.19) & 0.053 (0.16) &  0.015 (0.15) \\ 
 		0.9 & 0.78 & 0.013 (0.33) & 0.021 (0.41) & 0.033 (0.32) &  0.008 (0.33) \\ 
 		\hline \\[-1.8ex] 
 	\end{tabular} 
 \end{table}

\begin{table}[!htbp] \centering 
	\caption{Median bias (median absolute deviation) of the residual variance $\sigma^2_e$ in the surrogate model, $K=20$, average proportion of rejection 50\%.  Full stands for the model with fully available data, Cca for complete case analysis, Mean for the sample average imputation of missing $\rho_w$, and Beta for the stochastic multiple imputation procedure for missing $\rho_w$.} 
	\label{tab:surrog_sigmae1} 
	\begin{tabular}{@{\extracolsep{5pt}} cccccc} 
		\\[-1.8ex]\hline 
		\hline \\[-1.8ex] 
		$\rho_b$ & $\mathbb{E}[\rho_w]$ & Full & Cca & Mean  & Beta \\
		\hline \\[-1.8ex] 
		-0.4 & 0 & -0.177 (0.19) & -0.375 (0.24) & -0.169 (0.19) &  -0.172 (0.20) \\ 
		-0.4 & 0.25 & -0.191 (0.15) & -0.395 (0.18) & -0.164 (0.16) &  -0.187 (0.16) \\ 
		-0.4 & 0.50 & -0.197 (0.10) & -0.391 (0.12) & -0.145 (0.10) & -0.189 (0.10) \\ 
		-0.4 & 0.78 & -0.194 (0.05) & -0.415 (0.06) & -0.108 (0.05) & -0.187 (0.05) \\ 
		0 & 0 & -0.205 (0.26) & -0.406 (0.38) & -0.203 (0.25) &  -0.204 (0.26) \\ 
		0 & 0.25 & -0.201 (0.21) & -0.415 (0.26) & -0.203 (0.21) & -0.199 (0.21) \\ 
		0 & 0.50 & -0.193 (0.24) & -0.405 (0.29) & -0.192 (0.24) &  -0.190 (0.24) \\ 
		0 & 0.78 & -0.188 (0.20) & -0.392 (0.23) & -0.188 (0.20) &  -0.186 (0.20) \\ 
		0.4 & 0 & -0.171 (0.14) & -0.380 (0.17) & -0.177 (0.14) &  -0.168 (0.14) \\ 
		0.4 & 0.25 & -0.170 (0.08) & -0.375 (0.09) & -0.199 (0.09)  & -0.166 (0.08) \\ 
		0.4 & 0.50 & -0.162 (0.24) & -0.347 (0.29) & -0.204 (0.23)  & -0.159 (0.23) \\ 
		0.4 & 0.78 & -0.148 (0.20) & -0.322 (0.24) & -0.216 (0.20)  & -0.146 (0.20) \\ 
		0.7 & 0 & -0.135 (0.16) & -0.298 (0.20) & -0.144 (0.16) &  -0.130 (0.16) \\ 
		0.7 & 0.25 & -0.126 (0.11) & -0.282 (0.14) & -0.173 (0.12)  & -0.121 (0.11) \\ 
		0.7 & 0.50 & -0.112 (0.07) & -0.252 (0.09) & -0.183 (0.07)  & -0.106 (0.07) \\ 
		0.7 & 0.78 & -0.097 (0.19) & -0.209 (0.27) & -0.176 (0.19)  & -0.095 (0.20) \\ 
		0.9 & 0 & -0.104 (0.20) & -0.190 (0.24) & -0.119 (0.20) &  -0.095 (0.20) \\ 
		0.9 & 0.25 & -0.085 (0.23) & -0.190 (0.28) & -0.137 (0.23)  & -0.072 (0.23) \\ 
		0.9 & 0.50 & -0.066 (0.21) & -0.163 (0.25) & -0.134 (0.21) & -0.052 (0.21) \\ 
		0.9 & 0.78 & -0.048 (0.16) & -0.108 (0.19) & -0.103 (0.16) & -0.037 (0.16) \\ 
		\hline \\[-1.8ex] 
	\end{tabular} 
\end{table}

\begin{table}[!htbp] \centering 
	\caption{Median bias (median absolute deviation) of the residual variance $\sigma^2_e$ in the surrogate model, $K=30$, average proportion of rejection 50\%  Full stands for the model with fully available data, Cca for complete case analysis, Mean for the sample average imputation of missing $\rho_w$, and Beta for the stochastic multiple imputation procedure for missing $\rho_w$.} 
	\label{tab:surrog_sigmae2} 
	\begin{tabular}{@{\extracolsep{5pt}} cccccc} 
		\\[-1.8ex]\hline 
		\hline \\[-1.8ex] 
		$\rho_b$ & $\mathbb{E}[\rho_w]$ & Full & Cca & Mean  & Beta \\ 
		\hline \\[-1.8ex] 
		-0.4 & 0 & -0.120 (0.10) & -0.247 (0.06) & -0.115 (0.08) &  -0.116 (0.10) \\ 
		-0.4 & 0.25 & -0.123 (0.23) & -0.243 (0.28) & -0.093 (0.23) & -0.118 (0.23) \\ 
		-0.4 & 0.50 & -0.124 (0.20) & -0.262 (0.23) & -0.067 (0.20) & -0.118 (0.20) \\ 
		-0.4 & 0.78 & -0.130 (0.17) & -0.270 (0.21) & -0.046 (0.17)  & -0.120 (0.17) \\ 
		0 & 0 & -0.130 (0.13) & -0.272 (0.16) & -0.131 (0.14) &  -0.127 (0.14) \\ 
		0 & 0.25 & -0.129 (0.09) & -0.272 (0.11) & -0.131 (0.09) &  -0.127 (0.09) \\ 
		0 & 0.50 & -0.124 (0.15) & -0.271 (0.21) & -0.127 (0.16) &  -0.122 (0.15) \\ 
		0 & 0.78 & -0.132 (0.20) & -0.261 (0.23) & -0.124 (0.19) &  -0.128 (0.20) \\ 
		0.4 & 0 & -0.114 (0.24) & -0.244 (0.28) & -0.121 (0.24) &  -0.109 (0.24) \\ 
		0.4 & 0.25 & -0.112 (0.22) & -0.234 (0.27) & -0.143 (0.22) &  -0.110 (0.22) \\ 
		0.4 & 0.50 & -0.104 (0.18) & -0.225 (0.22) & -0.149 (0.18) &  -0.097 (0.18) \\ 
		0.4 & 0.78 & -0.093 (0.09) & -0.211 (0.04) & -0.158 (0.07) &  -0.089 (0.09) \\ 
		0.7 & 0 & -0.095 (0.21) & -0.200 (0.25) & -0.107 (0.21) &  -0.086 (0.21) \\ 
		0.7 & 0.25 & -0.086 (0.20) & -0.186 (0.24) & -0.135 (0.20) &  -0.080 (0.20) \\ 
		0.7 & 0.50 & -0.072 (0.18) & -0.158 (0.22) & -0.138 (0.18) &  -0.059 (0.18) \\ 
		0.7 & 0.78 & -0.065 (0.15) & -0.134 (0.18) & -0.146 (0.15) &  -0.059 (0.15) \\ 
		0.9 & 0 & -0.069 (0.11) & -0.153 (0.09) & -0.085 (0.11) &  -0.056 (0.10) \\ 
		0.9 & 0.25 & -0.058 (0.10) & -0.129 (0.14) & -0.105 (0.11)  & -0.046 (0.10) \\ 
		0.9 & 0.50 & -0.044 (0.19) & -0.098 (0.22) & -0.105 (0.19)  & -0.025 (0.19) \\ 
		0.9 & 0.78 & -0.031 (0.05) & -0.068 (0.07) & -0.084 (0.05)  & -0.019 (0.05) \\ 
		\hline \\[-1.8ex] 
	\end{tabular} 
\end{table}

\begin{table}[!htbp] \centering 
	\caption{Median bias (median absolute deviation) of the residual variance $\sigma^2_e$ in the surrogate model, $K=20$, average proportion of rejection 30\%.  Full stands for the model with fully available data, Cca for complete case analysis, Mean for the sample average imputation of missing $\rho_w$, and Beta for the stochastic multiple imputation procedure for missing $\rho_w$.} 
	\label{tab:surrog_sigmae3} 
	\begin{tabular}{@{\extracolsep{5pt}} cccccc} 
		\\[-1.8ex]\hline 
		\hline \\[-1.8ex] 
		$\rho_b$ & $\mathbb{E}[\rho_w]$ & Full & Cca & Mean & Beta \\
		\hline \\[-1.8ex] 
		-0.4 & 0 & -0.176 (0.24) & -0.264 (0.34) & -0.171 (0.23) &  -0.173 (0.24) \\ 
		-0.4 & 0.25 & -0.190 (0.21) & -0.276 (0.30) & -0.170 (0.20) & -0.188 (0.21) \\ 
		-0.4 & 0.50 & -0.197 (0.24) & -0.288 (0.33) & -0.159 (0.24) & -0.192 (0.24) \\ 
		-0.4 & 0.78 & -0.194 (0.20) & -0.292 (0.28) & -0.137 (0.20) &  -0.191 (0.20) \\ 
		0 & 0 & -0.203 (0.14) & -0.301 (0.21) & -0.203 (0.15) &  -0.201 (0.14) \\ 
		0 & 0.25 & -0.202 (0.08) & -0.308 (0.03) & -0.201 (0.06) &  -0.200 (0.09) \\ 
		0 & 0.50 & -0.194 (0.24) & -0.292 (0.34) & -0.189 (0.23) &  -0.191 (0.24) \\ 
		0 & 0.78 & -0.189 (0.20) & -0.294 (0.28) & -0.185 (0.20) &  -0.188 (0.20) \\ 
		0.4 & 0 & -0.172 (0.16) & -0.259 (0.23) & -0.180 (0.16) &  -0.172 (0.16) \\ 
		0.4 & 0.25 & -0.172 (0.11) & -0.254 (0.17) & -0.187 (0.12) &  -0.167 (0.12) \\ 
		0.4 & 0.50 & -0.164 (0.07) & -0.240 (0.09) & -0.187 (0.07) & -0.160 (0.07) \\ 
		0.4 & 0.78 & -0.147 (0.19) & -0.226 (0.26) & -0.183 (0.19) &  -0.146 (0.19) \\ 
		0.7 & 0 & -0.136 (0.20) & -0.199 (0.28) & -0.142 (0.20) & -0.131 (0.20) \\ 
		0.7 & 0.25 & -0.127 (0.23) & -0.194 (0.33) & -0.156 (0.24) &  -0.122 (0.24) \\ 
		0.7 & 0.50 & -0.112 (0.21) & -0.171 (0.30) & -0.155 (0.21) &  -0.108 (0.21) \\ 
		0.7 & 0.78 & -0.097 (0.16) & -0.147 (0.23) & -0.144 (0.16) & -0.097 (0.16) \\ 
		0.9 & 0 & -0.105 (0.11) & -0.155 (0.00) & -0.116 (0.05) &  -0.098 (0.10) \\ 
		0.9 & 0.25 & -0.085 (0.23) & -0.129 (0.32) & -0.112 (0.23) &  -0.078 (0.23) \\ 
		0.9 & 0.50 & -0.066 (0.20) & -0.101 (0.28) & -0.103 (0.20) &  -0.059 (0.20) \\ 
		0.9 & 0.78 & -0.048 (0.17) & -0.071 (0.25) & -0.076 (0.17) &  -0.040 (0.17) \\ 
		\hline \\[-1.8ex] 
	\end{tabular} 
\end{table}

\begin{table}[!htbp] \centering 
	\caption{Median bias (median absolute deviation) of the residual variance $\sigma^2_e$ in the surrogate model, $K=30$, average proportion of rejection 30\%.  Full stands for the model with fully available data, Cca for complete case analysis, Mean for the sample average imputation of missing $\rho_w$, and Beta for the stochastic multiple imputation procedure for missing $\rho_w$.} 
	\label{tab:surrog_sigmae4} 
	\begin{tabular}{@{\extracolsep{5pt}} cccccc} 
		\\[-1.8ex]\hline 
		\hline \\[-1.8ex] 
		$\rho_b$ & $\mathbb{E}[\rho_w]$ & Full & Cca & Mean  & Beta \\
		\hline \\[-1.8ex] 
		-0.4 & 0 & -0.120 (0.13) & -0.166 (0.19) & -0.115 (0.14) &  -0.119 (0.13) \\ 
		-0.4 & 0.25 & -0.123 (0.09) & -0.173 (0.06) & -0.099 (0.08) &  -0.119 (0.09) \\ 
		-0.4 & 0.50 & -0.124 (0.12) & -0.177 (0.17) & -0.087 (0.12) &  -0.119 (0.12) \\ 
		-0.4 & 0.78 & -0.130 (0.20) & -0.193 (0.26) & -0.074 (0.19) & -0.126 (0.20) \\ 
		0 & 0 & -0.130 (0.24) & -0.178 (0.33) & -0.129 (0.24) & -0.130 (0.24) \\ 
		0 & 0.25 & -0.129 (0.22) & -0.185 (0.32) & -0.128 (0.22) &  -0.128 (0.22) \\ 
		0 & 0.50 & -0.124 (0.18) & -0.182 (0.21) & -0.124 (0.18) &  -0.123 (0.18) \\ 
		0 & 0.78 & -0.132 (0.09) & -0.179 (0.00) & -0.119 (0.07) &  -0.128 (0.09) \\ 
		0.4 & 0 & -0.114 (0.21) & -0.163 (0.30) & -0.120 (0.21) & -0.113 (0.21) \\ 
		0.4 & 0.25 & -0.112 (0.20) & -0.160 (0.28) & -0.131 (0.20)  & -0.112 (0.20) \\ 
		0.4 & 0.50 & -0.104 (0.18) & -0.154 (0.26) & -0.129 (0.19)  & -0.100 (0.19) \\ 
		0.4 & 0.78 & -0.093 (0.15) & -0.149 (0.22) & -0.128 (0.15)  & -0.091 (0.15) \\ 
		0.7 & 0 & -0.095 (0.11) & -0.137 (0.04) & -0.103 (0.10) &  -0.090 (0.11) \\ 
		0.7 & 0.25 & -0.086 (0.06) & -0.124 (0.08) & -0.115 (0.06) &  -0.081 (0.06) \\ 
		0.7 & 0.50 & -0.072 (0.19) & -0.110 (0.25) & -0.113 (0.19) &  -0.065 (0.19) \\ 
		0.7 & 0.78 & -0.065 (0.24) & -0.097 (0.28) & -0.107 (0.23) &  -0.061 (0.24) \\ 
		0.9 & 0 & -0.069 (0.19) & -0.103 (0.22) & -0.081 (0.19) & -0.062 (0.19) \\ 
		0.9 & 0.25 & -0.058 (0.12) & -0.082 (0.14) & -0.086 (0.12)  & -0.050 (0.12) \\ 
		0.9 & 0.50 & -0.044 (0.06) & -0.068 (0.07) & -0.080 (0.06)  & -0.033 (0.06) \\ 
		0.9 & 0.78 & -0.031 (0.26) & -0.047 (0.31) & -0.058 (0.26)  & -0.024 (0.26) \\ 
		\hline \\[-1.8ex] 
	\end{tabular} 
\end{table} 
%\end{document}

\end{document}